\documentclass{svmult}

\usepackage[utf8]{inputenc}
\usepackage{xcolor}
\usepackage{hyperref}
\hypersetup{
    colorlinks,
    linkcolor={red!50!black},
    citecolor={blue!50!black},
    urlcolor={blue!80!black}
}
\usepackage{amsmath,amsfonts,amssymb}
\usepackage{cases}
\usepackage{graphicx}
\usepackage[caption=false,font=small]{subfig}
\usepackage{booktabs,diagbox}
\usepackage{multirow}
\usepackage{enumerate}
\usepackage{paralist}
\usepackage{acro}
\usepackage{cleveref}
\usepackage[textsize=tiny]{todonotes}
\usepackage{color}

\usepackage{localmacros}
\usepackage{paperacronyms}

\crefname{equation}{}{}
\crefname{subequation}{}{}

\title*{Image reconstruction in dynamic inverse problems with temporal models}
\titlerunning{Image reconstruction in dynamic inverse problems}
\author{Andreas Hauptmann and Ozan Öktem and Carola Schönlieb}
\institute{Andreas Hauptmann \at University of Oulu, Research Unit of Mathematical Sciences, Oulu, Finland, 
\at University College London, Department of Computer Science, London, United Kingdom,
\email{andreas.hauptmann@oulu.fi}
\and Ozan Öktem \at Department of Mathematics, KTH - Royal Institute of Technology, \email{ozan@kth.se}
\and Carola-Bibiane Schönlieb \at Department of Applied Mathematics and Theoretical Physics, University of Cambridge, \email{cbs31@cam.ac.uk}}

\begin{document}
\maketitle

\abstract{The paper surveys variational approaches for image reconstruction in dynamic inverse problems.
Emphasis is on methods that rely on parametrised temporal models. These are here encoded as diffeomorphic deformations with time dependent parameters, or as motion constrained reconstruction where the motion model is given by a partial differential equation. The survey also includes recent development in integrating deep learning for solving these computationally demanding variational methods. Examples are given for 2D dynamic tomography, but methods apply to general inverse problems.}

\section{Introduction} 
Dynamic inverse problems in imaging refers to the case when the object being imaged undergoes a temporal evolution during the data acquisition.
The resulting data in such an inverse problem is a time (or quasi-time) series and due to limited sampling speed typically highly under-sampled. Failing to account for the dynamic nature of the imaged object will lead to severe degradation in image quality and hence there is a strong need for advanced modelling of the involved dynamics by incorporating temporal models in the reconstruction task.

The need for dynamic imaging arises for instance in various tomographic imaging studies in medicine, such as imaging moving organs (respiratory and cardiac motion) with \ac{CT} \cite{Kwong:2015aa}, \ac{PET} or \ac{MRI} \cite{lustig2006kt} and in functional imaging studies by means of dynamic \ac{PET} \cite{Rahmim:2019aa} or functional MRI \cite{glover2011overview}. In functional imaging studies, the dynamic information is crucial for the diagnostic value to asses functionality of organs or tracking an injected tracer. Spatiotemporal imaging also arises in life sciences \cite{Mokso:2014aa} where it is crucial to understand dynamics and interactions of organisms. Lastly, applications in material sciences \cite{Ruhlandt:2017aa,De-Schryver:2018aa} and process monitoring \cite{chen2018extended} rely on the capabilities dynamic image reconstruction.

Mathematically, solving dynamic inverse problems in imaging, or spatiotemporal image reconstruction, aims to recover a time dependent image from a measured time-series. Since the measured time series is typically highly undersampled in each time instance, the reconstruction task is ill-posed and additional prior knowledge is needed to recover a meaningful spatiotemporal image. One such prior assumption can be made on the type of dynamics in the studied object, which can regularise the reconstruction task by penalising unrealistic motion. 

This review focuses on methods that recover the tomographic image \emph{jointly} with determining parameters in a temporal model that regulates its time evolution. We discuss how temporal models can be incorporated into a variational framework and present two primary choices to incorporate temporal information. Either by deforming a static template using time dependent parameter or constrain the variational formulation using an explicit motion model. 

\section{Spatiotemporal inverse problems}\label{sec:overview}
The starting point is to mathematically formalise the notion of a spatiotemporal inverse problem, which refers to the task of recovering a time dependent image from (time dependent) noisy indirect observations.
\begin{description}
\item[Image:]
The time dependent image is formally represented by a function $\signal \colon [0,T] \times \signaldomain \to \Real^k$ where $k$ is the number of image channels ($k=1$ for greyscale images) and $\signaldomain \subset \Real^d$ is the image domain. 

We henceforth assume $\signal(t,\Cdot) \in \RecSpace$ where $\RecSpace$ (reconstruction space) is some vector space of $\Real^k$-valued functions on $\signaldomain \subset \Real^d$ that, unless otherwise stated, is a Hilbert space under the $\LpSpace^2$-inner product. 
\\[0.1em]
\item[Data:]
Data is represented by a time dependent function $\data \colon [0,T] \times \datadomain \to \Real^l$ 
where $\datadomain$ is some manifold that is defined by the acquisition geometry and $l$ is the number of data channels. 
Likewise, we assume that $\data(t,\Cdot) \in \DataSpace$ where $\DataSpace$ (data space) is some vector space of $\Real^l$-valued functions on $\datadomain$ that, unless otherwise stated, is a Hilbert space under the $\LpSpace^2$-inner product. 
Actual measured data represents a digitisation of this function by sampling on $[0,T] \times \datadomain$.
\\[0.1em]
\item[Spatiotemporal inverse problem:]
This is the task of recovering a temporal image $t \mapsto \signal(t,\Cdot) \in \RecSpace$ from time series data $t \mapsto \data(t,\Cdot) \in\DataSpace$ where 
\begin{equation}\label{eq:InvProb}  
     \data(t,\Cdot) = \ForwardOp\bigl(t, \signal(t,\Cdot)\bigr)(t,\Cdot) + \datanoise(t,\Cdot) \quad \text{on $\datadomain$ for $t \in [0,T]$.}
\end{equation}
Here, $\ForwardOp(t, \Cdot) \colon \RecSpace \to \DataSpace$ is a (possibly time-dependent) forward operator that models how an image at time $t$ gives rise to data in absence of noise or measurement errors.
The observation noise in data is accounted for by $\datanoise(t,\Cdot) \in \DataSpace$, which can be seen as a single random realisation of a $\DataSpace$-valued random variable that models measurement noise.
\end{description}
\begin{remark}
  The formulation in \cref{eq:InvProb} also covers cases when noise in data depends on the signal strength, like Poisson noise. Simply assume $\datanoise(t,\Cdot)$ in \cref{eq:InvProb} is a sample of the random variable $\stdatanoise(t,\Cdot) := \stdata(t,\Cdot) - \ForwardOp\bigl(t, \signal(t,\Cdot)\bigr)$ where $\stdata(t,\Cdot)$ is the $\DataSpace$-valued random variable generating data.
\end{remark}

Special cases of \cref{eq:InvProb} arise depending on how the time dependency enters into the problem.
In particular, the following three components can depend on time independently of each other:
\begin{enumerate}[(a)]
\item Forward operator: The forward model may depend intrinsically on time.
\item Data acquisition geometry: The way the forward operator is sampled has a specific time dependency.
\item Image: The image to be recovered depends on time.
\end{enumerate}
Next, an important special case is when data in \cref{eq:InvProb} is observed at discrete time instances $0 \leq t_0 < \ldots < t_n \leq T$. 
Then, \cref{eq:InvProb} reduces to the task of recovering images $\signal_j \in \RecSpace$ from data $\data_j \in \DataSpace$ where 
\begin{equation}\label{eq:InvProbTimeDiscr}  
     \data_j = \ForwardOp_j(\signal_j) + \datanoise_j 
     \quad \text{for $j =1,\ldots, n$.}
\end{equation}
In the above, we have made use of the following notation for $j=1,\ldots,n$: 
\begin{equation}\label{eq:NotationTimeDiscr}
\begin{alignedat}{2}
  \data_j &:= \data(t_j,\Cdot) \in \DataSpace &\qquad \signal_j &:= \signal(t_j,\Cdot) \in \RecSpace \\
  \datanoise_j &:= \datanoise(t_j,\Cdot) \in \DataSpace
    &\qquad \ForwardOp_j &:= \ForwardOp\bigl(t_j, \Cdot) \colon \RecSpace \to \DataSpace.
\end{alignedat}
\end{equation}

\subsection{Reconstruction without explicit temporal models}\label{sec:tempindependentrecon}
The inverse problem in \cref{eq:InvProb} is almost always ill-posed, so solving it requires regularisation both regarding the spatial and temporal variation of the image. 
A variational approach for reconstructing the image trajectory $t \mapsto \signal(t,\Cdot)$ that does not use any explicit temporal model reads as 
\begin{equation}\label{eq:DirectSpatioTempReg}
\argmin_{t \mapsto \signal(t,\Cdot) \in \RecSpace} \int_0^T 
  \Bigl[ \DataDiscrep\Bigl(\ForwardOp\bigl(t,\signal(t,\Cdot)\bigr),\data(t,\Cdot)\Bigr)  
    + \RegFunc_{\deforparam}\bigl(t,\signal(t,\Cdot)\bigr)
  \Bigr] \,dt.
\end{equation}
Here, $\DataDiscrep \colon \DataSpace \times \DataSpace \to \Real$ is the data fidelity term (data-fit), which is ideally chosen as an appropriate affine transform of the negative log-likelihood of data \cite{Bertero:2008aa}.
The term $\RegFunc_{\deforparam} \colon \RecSpace \to \Real$ is a parametrised regulariser that accounts for a priori knowledge about the image. It is common to separately regularise the spatial and temporal components, e.g., by considering  
\[  \RegFunc_{\deforparam}\bigl(t,\signal(t,\Cdot)\bigr) := 
        \RegFuncSpat_{\gamma}\bigl(\signal(t,\Cdot)\bigr)
       +
        \RegFuncTemp_{\tau}\bigl(\partial_t \signal(t,\Cdot)\bigr) 
\quad\text{for $\deforparam = (\gamma,\tau)$.}
\]

In the above, $\RegFuncSpat_{\gamma} \colon \RecSpace \to \Real$ is a spatial regulariser and $\RegFuncTemp_{\tau} \colon \RecSpace \to \Real$ is a temporal regulariser.
The spatial regulariser is commonly of the form $\RegFuncSpat_{\gamma} := \gamma \RegFuncSpat$ where $\gamma >0$ and $\RegFuncSpat \colon \RecSpace \to \Real$ is some `energy' functional.
There is a well-developed theory for how to choose the latter in order to promote solutions of an inverse problem with specific type of regularity, e.g., a suitable choice for $\mathcal{H}^1(\signaldomain)$-regularity is
\begin{equation}\label{eq:TikReg}
    \RegFuncSpat(\signal) := \int_{\signaldomain} \bigl\vert \nabla \signal(x) \bigr\vert^2 \der x.
\end{equation}
On the other hand, if the image has edges that need to be preserved, then $\BVSpace(\signaldomain)$-regularity 
is more natural and a \ac{TV}-regulariser is a better choice \cite{RuOs92}.
This regulariser is for $\signal\in W^{1,1}(\signaldomain)$ expressible as 
\begin{equation}\label{eq:TVReg}
 \RegFuncSpat(\signal) := \int_{\signaldomain} \bigl\vert \nabla \signal(x) \bigr\vert \der x.
\end{equation}
Other choices may include higher order terms to the total variation functional, like in total generalised variation, see \cite{Benning:2018aa,Scherzer:2009aa} for a survey.

The choice of temporal regulariser is much less explored. This functional accounts for a priori temporal regularity. Similarly to \cref{eq:TikReg} one can here think of a smoothness prior for slowly evolving images 
\begin{equation}\label{eq:TikRegTime}
    \RegFuncTemp(\partial_t\signal) := \int_{\signaldomain} \bigl\vert \partial_t\signal(x) \bigr\vert^2 \der x,
\end{equation}
or a total variation type of penalty for changes that are small or occur step-wise (image changes step-wise). 
The regulariser \cref{eq:TikRegTime} acts point-wise in time and full temporal dependency is obtained by integrating over time in \cref{eq:DirectSpatioTempReg}.

Methods for solving \cref{eq:InvProb} based on \cref{eq:DirectSpatioTempReg} can be used when there is no explicit temporal model that connects images and data across time. 
Hence, the are applicable to a wide range of dynamic inverse problems as outlined in \cite{Schmitt2002,Schmitt2002a}.
More specific imaging related applications are \cite{feng2014golden,lustig2006kt,steeden2018real} for spatiotemporal compressed sensing in dynamic \ac{MRI}. Here, the temporal regularity is enforced by a sparsifying transform (or total variation). Further examples are $\mu$\ac{CT} imaging of dynamic processes \cite{bubba2017shearlet,niemi2015dynamic} and process monitoring with electrical resistance tomography \cite{chen2018extended}.

\begin{remark}
When data is time discretised, then one also has the option to consider reconstructing images at each time step independently.
An example of this is to recover the image at $t_j$ by using a variational regularisation method, i.e., as $\signal_j \approx \widehat{\signal}_j$ where 
\begin{equation}\label{eq:SpatioTempRecIndep}
  \widehat{\signal}_j := \argmin_{\signal \in \RecSpace} 
     \Bigl\{ \DataDiscrep\bigl(\ForwardOp_j(\signal),\data_j \bigr) 
      + \RegFuncSpat_{\gamma_j}(\signal) \Bigr\}
  \quad\text{for $j=1,\ldots, n$.}
\end{equation}
\end{remark}

\emph{Our emphasis will henceforth be on methods for solving \cref{eq:InvProb} that utilise more explicit temporal models.} 

\subsection{Reconstruction using a motion model}\label{sec:MotionRec}
The idea here is to assume that a solution $t \mapsto \signal(t,\Cdot) \in \RecSpace$ to \cref{eq:InvProb} has a time evolution that can be modelled by a \emph{motion model}.
Restating this assumption mathematically, we assume there is an operator $\MotionOp \colon [0,T] \times \RecSpace \to \RecSpace$ (motion model) such that 
\begin{equation}\label{eq:MotionModel}
  \MotionOp\bigl( t, \signal(t,\Cdot) \bigr) = 0 \quad\text{on $\signaldomain$ whenever $t \mapsto \signal(t,\Cdot)$ solves \cref{eq:InvProb}.}
\end{equation}
Hence, \cref{eq:InvProb} can be re-phrased as the task of recovering the image trajectory $t \mapsto \signal(t,\Cdot) \in \RecSpace$ along with its motion model $\MotionOp \colon [0,T] \times \RecSpace \to \RecSpace$ from time series data $t \mapsto \data(t,\Cdot) \in\DataSpace$ where 
\begin{equation}\label{eq:InvProbMotionModel}  
  \begin{split}
    & \data(t,\Cdot) = \ForwardOp\bigl(t, \signal(t,\Cdot)\bigr)(t,\Cdot) + \datanoise(t,\Cdot) \text{ on $\datadomain$} \\
    & \text{ s.t. $\MotionOp\bigl( t, \signal(t,\Cdot) \bigr) = 0$ on $\signaldomain$.}
  \end{split}    
  \quad\text{for $t \in [0,T]$.}
\end{equation}

\subsubsection{Parametrised motion models}
An important special case is when the motion model only depends on time through a time dependent parameter, i.e., there is $\MotionOp_{\motionparam} \colon \RecSpace \to \RecSpace$ for $\motionparam \in \MotionParamSet$ such that 
\begin{equation}\label{eq:MotionModelParam}
  \MotionOp_{\motionparam_t}\bigl( \signal(t,\Cdot) \bigr) = 0 
  \quad\text{on $\signaldomain$ whenever $t \mapsto \signal(t,\Cdot)$ solves \cref{eq:InvProb},}
\end{equation}
for some $t \mapsto \motionparam_t$. 
Then, \cref{eq:InvProb} can be re-phrased as the task to recover $t \mapsto \signal(t,\Cdot) \in \RecSpace$ along with motion parameter $t \mapsto \motionparam_t \in \MotionParamSet$ from time series data $t \mapsto \data(t,\Cdot) \in\DataSpace$ where 
\begin{equation}\label{eq:InvProbMotionModelParam}  
  \begin{split}
    & \data(t,\Cdot) = \ForwardOp\bigl(t, \signal(t,\Cdot)\bigr)(t,\Cdot) + \datanoise(t,\Cdot) \text{ on $\datadomain$} \\
    & \text{ s.t. $\MotionOp_{\motionparam_t}\bigl(\signal(t,\Cdot) \bigr) = 0$ on $\signaldomain$.}
  \end{split}    
  \quad\text{for $t \in [0,T]$.}
\end{equation}

The assumption in \cref{eq:MotionModelParam} may act as a regularisation since it introduces a model for how images vary across time. 
In particular, the inverse problem in \cref{eq:InvProbMotionModelParam} is challenging, but still easier to handle than the one in \cref{eq:InvProb}.
However, solving \cref{eq:InvProbMotionModelParam} will still most likely require regularisation. 
Approaches surveyed in \cref{sec:PDE} represent different ways for doing this based on the setting where $\MotionOp_{\motionparam}\colon \RecSpace \to \RecSpace$ is given as a differential operator (involving differentiation in both temporal and spatial variables).
Then parameter set $\MotionParamSet$ is a vector space of vector fields $\motionparam \colon \signaldomain \to \Real^d$ with sufficient regularity, so $\motionparam_t$ corresponds to a velocity field.
With these assumptions, \cref{eq:MotionModelParam} is a differential equation that constrains the temporal evolution of the solution to \cref{eq:InvProb} and \cref{eq:InvProbMotionModelParam} corresponds to reconstructing the image jointly with its motion model.

\subsubsection{General variational formulation}\label{sec:MotionConstrained}
It is quite natural to adopt a variational approach for solving \cref{eq:InvProbMotionModelParam}. 
In fact, many of the state-of-the-art methods are of the form 
\begin{equation}\label{eq:SpatioTempConstraintMotModel}
\begin{split}
  &\argmin_{\substack{\signal(t,\Cdot) \in \RecSpace \\ \deforparam_t \in \DeforParamSet}} 
  \biggl\{ \int_0^T \Bigl[ \DataDiscrep\Bigl(\ForwardOp\bigl(t, \signal(t,\cdot) \bigr),\data(t,\Cdot)\Bigr) 
    + 	 \RegFuncTemp_\tau(t,\deforparam_t) +  \RegFuncSpat_\gamma(\signal(t,\cdot)) \Bigr] \,dt
  \biggr\}. \\
  & \text{s.t. } \Psi_{\deforparam_t}\bigl(\signal(t,\cdot)\bigr) = 0, \quad \text{for $t\in [0,T]$.}
  \end{split}
\end{equation}
Just as for \cref{eq:DirectSpatioTempReg}, one here needs to choose $\RegFuncSpat_{\gamma} \colon \RecSpace \to \Real$ (spatial regulariser) and $\RegFuncTemp_{\tau}(t, \Cdot) \colon \RecSpace \to \Real$ (temporal regulariser), whereas $\DataDiscrep \colon \DataSpace \times \DataSpace \to \Real$ is derived from a statistical model for the noise in data.

In practice, the hard constrained formulation might be too restrictive and we rather aim to solve a penalised version, where the motion constraint is incorporated as a regulariser, see \cref{sec:PDE} for further detials.
Next, for data that is time discretised, the formulation in \cref{eq:SpatioTempConstraintMotModel} reduces to a series of reconstruction and registration problems that are solved simultaneously. 
Practically, the optimisation is usually performed in an alternating way, where first a dynamic reconstruction $\signal(t,\cdot)$ for $t\in[0,T]$ is obtained, followed by an update of the motion parameters $t \mapsto \deforparam_t$. This procedure is then iterated until a sufficient convergence criterium is fulfilled \cite{Burger:2018aa}. Interpreted in a Bayesian setting, this approaches compares to smoothing, see for instance \cite{Burger:2017aa} for a discussion on this topic.

\subsection{Reconstruction using a deformable template}\label{sec:DeforTemp}
The idea here is that when solving \cref{eq:InvProb}, the temporal model for $t \mapsto \signal(t,\Cdot) \in \RecSpace$ is given by deforming a fixed (time independent) template $\template \in \RecSpace$ using a time dependent parametrisation of a deformation operator.

\subsubsection{Deformation operators}\label{sec:DeforTempInvProb}
To formalise the underlying assumption in reconstruction with a deformable template, we assume there is a fixed family $\{ \DeforOp_{\deforparam} \}_{\deforparam \in \DeforParamSet}$ of mappings (deformation operators) 
\begin{equation}\label{eq:DefOpParam}
 \DeforOp_{\deforparam} \colon \RecSpace \to \RecSpace
   \quad\text{for $\deforparam \in \DeforParamSet$.}
\end{equation}
Next, we assume that   
\begin{equation}\label{eq:DeforTemp}
 \signal(t,\Cdot) = \DeforOp_{\deforparam_t}(\template)
   \quad\text{on $\signaldomain$ whenever $t \mapsto \signal(t,\Cdot)$ solves \cref{eq:InvProb},}
\end{equation}
for some $t \mapsto \deforparam_t \in \DeforParamSet$ and $\template \in \RecSpace$.
Then, \cref{eq:InvProb} can be re-phrased as the inverse problem of recovering $\template \in \RecSpace$ and $t \mapsto \deforparam_t \in \DeforParamSet$ from time series data $\data(t,\Cdot) \in\DataSpace$ where 
\begin{equation}\label{eq:InvProbDeforTemp}  
     \data(t,\Cdot) = \ForwardOp\bigl(t, \DeforOp_{\deforparam_t}(\template) \bigr) + \datanoise(t,\Cdot) 
     \quad \text{on $\datadomain$ for $t \in [0,T]$.}
\end{equation}

The assumption in \cref{eq:DeforTemp} may act as a regularisation since it introduces a model for how images vary across time. 
In particular, the inverse problem in \cref{eq:InvProbDeforTemp} is challenging, but still easier to handle than the one in \cref{eq:InvProb}.
However, solving \cref{{eq:InvProbDeforTemp}} will still most likely require regularisation. 
Variational approaches are suitable for this purpose, but these typically involve optimisation over the parameter set $\DeforParamSet$ so it is desirable to ensure $\DeforParamSet$ has a vector space structure. 
\Cref{sec:ODE} surveys different approaches for solving \cref{eq:InvProbDeforTemp} based on the setting where the deformation operator is a diffeomorphic deformation. 

\begin{remark}\label{rem:MotionVsDefor}
Comparing assumption \cref{eq:DeforTemp} with \cref{eq:MotionModel}, we see that they are equivalent if 
\[ \MotionOp\bigl(t,\DeforOp_{\deforparam_t}(\template) \bigr) = 0
    \quad\text{holds on $\signaldomain$ for $t \in [0,T]$.} 
\]
Hence, it is sometimes possible to view a motion model as deforming a template using a deformation operator with time dependent parametrisation.
Likewise, a deformation operator with a time dependent deformation acting on a template gives rise to a motion model.
\end{remark}

\subsubsection{General variational formulation}\label{sec:IndRegSpatioTempTimeContData}
A variational approach for solving \cref{eq:InvProbDeforTemp} can be formulated as  
\begin{equation}\label{eq:SpatioTempDeforTempReg}
  \argmin_{\substack{\template \in \RecSpace \\ t \mapsto \deforparam_t \in \DeforParamSet}} 
  \biggl\{ \int_0^T \Bigl[ \DataDiscrep\Bigl(\ForwardOp\bigl(t, \DeforOp_{\deforparam_t}(\template) \bigr),\data(t,\Cdot)\Bigr) 
    + \RegFuncTemp_{\tau}(t,\deforparam_t) \Bigr] \,dt +  \RegFuncSpat_{\gamma}(\template)
  \biggr\}.
\end{equation}
This is very similar to \cref{eq:DirectSpatioTempReg} with $\DataDiscrep \colon \DataSpace \times \DataSpace \to \Real$ denoting the data fidelity term and the regularisation term is a sum of a spatial and temporal regulariser:
\[
  \RegFuncSpat_{\gamma} \colon \RecSpace \to \Real
  \quad\text{and}\quad
  \RegFuncTemp_{\tau}(t,\Cdot) \colon \DeforParamSet \to \Real.
\]
The choice of the spatial regulariser $\RegFuncSpat_{\gamma}$ is a well-explored topic as outlined in \cref{sec:tempindependentrecon}.
In contrast, how to choose an appropriate temporal regulariser $\RegFuncTemp_{\tau}$ is less explored and closely linked to assumptions on $t \mapsto \deforparam_t$, which governs the time evolution of the image, see, e.g., \cref{sec:LDDMSpatioTemp} for an example.

\subsubsection{Time discretised data}\label{sec:IndRegSpatioTempTimeDiscData}
There are different strategies for solving \cref{eq:InvProbDeforTemp} when data is time discretised.
They differ depending on how the time discretised version is formulated, and in particular on how the initial template $\template$ is used for building up the images $\signal_j$ by means of a deformable templates model.
\begin{description}
\item[\textit{Independent trajectory:}]
The time discretised version of \cref{eq:InvProbDeforTemp} is formulated as the task of recovering $\template \in \RecSpace$ and $\deforparam_j \in \DeforParamSet$ from data $\data_j \in\DataSpace$ where 
\begin{equation}\label{eq:UnMatchedTrajectory}  
     \data_j = \ForwardOp_j\bigl(\DeforOp_{\deforparam_j}(\template) \bigr) + \datanoise_j
     \quad \text{for $j=1,\ldots, n$.}
\end{equation}
In the above, $\DeforOp_{\deforparam_j} \colon \RecSpace \to \RecSpace$ registers the initial template image $\template \in \RecSpace$ against a target image $\signal_j \in \RecSpace$ that is indirectly observed through data $\data_j \in \DataSpace$. In particular, the trajectory $t \mapsto \signal(t,\Cdot)$ is made up of images $\signal(t_j,\Cdot) = \signal_j := \DeforOp_{\deforparam_j}(\template)$ that are generated independently from each other by deforming the initial template $\template$.

One approach for solving \cref{eq:UnMatchedTrajectory} is to compute $\widehat{\signal}_j := \DeforOp_{\widehat{\deforparam}_{j}}(\widehat{\template})$ where 
\begin{multline}\label{eq:SpatioTempRecUnMatched}
 (\widehat{\template},\widehat{\deforparam}_1,\ldots,\widehat{\deforparam}_n) \in  \!\! \argmin_{\substack{\template \in \RecSpace \\ \deforparam_{1},\ldots, \deforparam_{n} \in \DeforParamSet}} 
  \biggl\{ \sum_{j=1}^n \Bigl[ 
    \DataDiscrep\Bigl(\ForwardOp_j\bigl(\DeforOp_{\deforparam_{j}}(\template) \bigr),\data_j \Bigr) 
\\[-1em]
    + \RegFuncTemp_{\tau}(\deforparam_j) + \RegFuncSpat_{\gamma}\bigl( \DeforOp_{\deforparam_{j}}(\template) \bigr) 
  \Bigr] 
  \biggr\}.
\end{multline}
Note that the choice of $\RegFuncTemp \colon \DeforParamSet \to \Real$ may introduce a dependency between $\widehat{\signal}_j$ and $\widehat{\signal}_{k}$ for $j\neq k$ even though $\signal_j$ and $\signal_{k}$ only depend on each other through the template $\template$. 
\item[\textit{Single trajectory:}]
Here the template $\template$ is only used once to generate the image at $t_1$, the sequence of images at $t_2, \ldots, t_n$ that make up the trajectory $t \mapsto \signal(t,\Cdot)$ are generated sequentially.
The time discretised version of \cref{eq:InvProbDeforTemp} now reduces to the task of recovering $\template \in \RecSpace$ and $\deforparam_j \in \DeforParamSet$ from data $\data_j \in\DataSpace$ where 
\begin{equation}\label{eq:SpatioTempRecMatchedSingle}  
     \data_j = \ForwardOp_j\bigl(\DeforOp_{\deforparam_j}(\signal_{j-1}) \bigr) + \datanoise_j
     \quad \text{for $j=1,\ldots, n$.}
\end{equation} 
In contrast to \cref{eq:UnMatchedTrajectory}, $\DeforOp_{\deforparam_j} \colon \RecSpace \to \RecSpace$ is used here to deform $\signal_{j-1} \in \RecSpace$ (image at time step $t_{j-1}$) to the target image $\signal_j \in \RecSpace$ that is indirectly observed through data $\data_j \in \DataSpace$. 
Note that one can re-write \cref{eq:SpatioTempRecMatchedSingle} as 
\begin{equation}\label{eq:SpatioTempRecMatchedSingle2}  
     \data_j = \ForwardOp_j\bigl((\DeforOp_{\deforparam_j} \circ \ldots \circ \DeforOp_{\deforparam_1})(\template) \bigr) + \datanoise_j
     \quad \text{for $j=1,\ldots, n$.}
\end{equation} 

One can attempt at solving \cref{eq:SpatioTempRecMatchedSingle} by the following intertwined scheme:
\begin{equation}\label{eq:SpatioTempRecMatchedTrajectory}
\begin{cases}
 \widehat{\template} = \displaystyle{\argmin_{\signal \in \RecSpace} }
     \Bigl\{ \DataDiscrep\Bigl(\ForwardOp_1(\signal),\data_1 \Bigr) 
      + \RegFunc(\signal) \Bigr\}
& \\[0.75em] 
  \widehat{\deforparam}_j \in \displaystyle{\argmin_{\deforparam \in \DeforParamSet}}
   \Bigl\{ \DataDiscrep\Bigl(\ForwardOp_j\bigl(\DeforOp_{\deforparam}(\widehat{\signal}_{j-1}) \bigr),\data_j \Bigr) 
& \\[0.75em] \qquad\qquad\qquad\qquad   
    + \RegFuncTemp_{\tau}(\deforparam) + \RegFuncSpat_{\gamma}\bigl( \DeforOp_{\deforparam}(\widehat{\signal}_{j-1}) \bigr) \Bigr\}
& \\  
   \widehat{\signal}_j := \DeforOp_{\widehat{\deforparam}_{j}}(\widehat{\signal}_{j-1})
\end{cases}
\quad\text{for $j=1,\ldots, n$.}
\end{equation}
Note that recursive time-stepping schemes of the above type can be related to filtering approaches in a Bayesian setting, see for instance  \cite{hakkarainen2019undersampled} for an application to dynamic X-ray tomography.
\end{description}

\section{Approaches based on \aclp{ODE}}\label{sec:ODE}
The reconstruction methods described here aim to solve \cref{eq:InvProbDeforTemp} using deformable templates (\cref{sec:DeforTemp}).

Images are elements in the Hilbert space $\RecSpace := L^2(\signaldomain, \Real)$ for some fixed bounded domain $\signaldomain \subset \Real^d$.
The deformation operator is given by acting with diffeomorphisms on images.
Hence, let $\Diff(\signaldomain)$ denote the group of diffeomorphisms (with composition as group law) and $(\diffeo,\template) \mapsto \diffeo.\template$ denotes the (group) action of $\Diff(\signaldomain)$ on $\RecSpace$.
In imaging there are now two natural options:
\begin{description}
\item[\textit{Geometric group action}:]
This group action simply moves image intensities without changing their grey-scale values, which corresponds to shape deformation:
\begin{equation}\label{eq:GeometricGroupAction}
   \diffeo.\template := \template \circ \diffeo^{-1} 
   \quad\text{for $\diffeo\in \Diff(\signaldomain)$ and $\template \in \RecSpace$.}
\end{equation}
\item[\textit{Mass preserving group action}:]
Image intensities are allowed to change, but one preserves the total mass:
\begin{equation}\label{eq:MassPreservingGroupAction}
   \diffeo.\template := \bigl\vert D \diffeo^{-1} \bigl\vert \, (\template \circ \diffeo^{-1})
   \quad\text{for $\diffeo\in \Diff(\signaldomain)$ and $\template \in \RecSpace$.}
\end{equation}
\end{description}
The second key component is to describe how the deformation operator is parametrised, which here becomes a parametrisation of the (sub)group of diffeomorphisms that are of interest.
Much of the theory is motivated by image registration and registation can in this setting be formulated as an optimisation over $\DeforParamSet$, so the chosen parametrisations is preferably an element in a vector space $\DeforParamSet$.

\subsection{Flow of diffeomorphisms and intensities}\label{sec:DiffeoGenerative}
In the \ac{LDDMM} framework for image registration, $\DeforParamSet = V$ where $V \subset C^{1}_0(\signaldomain,\Real^d)$ is a suitable Banach/Hilbert space of vector fields and the parametrised diffeomorphisms $G_V$ are obtained by considering solutions to \cref{eq:FlowEq}.

For a given velocity field $\vfield \colon [0,T]\times \signaldomain \to \signaldomain$, one can consider solutions to the flow equation 
\begin{equation}\label{eq:FlowEq}
\begin{cases}
  \dfrac{\der}{\der t} \diffeo(t,x) = \vfield\bigl(t,\diffeo(t,x)\bigr)  & \\[0.75em]
  \diffeo(0,x) = x &
\end{cases}  
\quad\text{for $x \in \signaldomain$ and $t \in [0,T]$.}
\end{equation}
Next, let $\VelocitySpace{1}{V}$ denote the vector space of mappings $\vfield \colon [0,T] \times \signaldomain \to \Real^d$ (velocity fields) where $\vfield(t,\Cdot) \in V$.
If $V$ is \emph{admissible}, then \cref{eq:FlowEq} has diffeomorphic solutions at any time $0 \leq t \leq 1$ whenever $\vfield \in \VelocitySpace{1}{V}$ (\cite[Theorem~7.11]{Younes:2019aa} and \cite{arguillere2014shape}).
Then, we can define $\diffeoflow{\vfield}{s,t} \colon \Real^d \to \Real^d$ as  
\begin{equation}\label{eq.FlowDiffeo}
    \diffeoflow{\vfield}{s,t} := \diffeo(t,\Cdot) \circ \diffeo(s,\Cdot)^{-1}
    \quad\text{for $s,t \in [0,T]$ and $\diffeo(t,\Cdot)$ solving \cref{eq:FlowEq}.}
\end{equation}
This is a diffeomorphism for any $0 \leq s,t \leq 1$, so $G_V$ defined below becomes a subgroup of diffeomorphisms parametrised by $V$: 
\begin{equation}\label{eq:GV}
 G_V := 
     \Bigl\{ \diffeo \colon \Real^d \to \Real^d : 
       \diffeo = \diffeoflow{\vfield}{0,T} \text{ for some $\vfield \in \VelocitySpace{1}{V}$} 
     \Bigr\}.
\end{equation}
\begin{remark}
$G_V$ is actually a subgroup of $\Diff^{1,\infty}_0(\signaldomain)$ \cite[Theorem~7.16]{Younes:2019aa} where
$\Diff^{p,\infty}_0(\signaldomain)$ is the group of $p$-diffeomorphisms that tend to the identity at infinity:
\[
  \Diff^{p,\infty}_0(\signaldomain) :=
    \bigl\{
       \diffeo \in \Diff^{p,\infty}(\signaldomain) :
       \diffeo - \Id \in C^p_0(\signaldomain,\Real^d)
    \bigr\}.
\]
Next, if $V$ is embedded in $C^p_0(\signaldomain,\Real^d)$, then $G_V$ is a subgroup of $\Diff^{p,\infty}_0(\signaldomain)$.
\end{remark}

Metamorphosis \cite[Chapter~13]{Younes:2019aa} is an extension of \ac{LDDMM} in the sense that it considers a flow equation that jointly evolves shape and intensities:
\begin{equation}\label{eq:MetaFlow}
\begin{cases}
  \dfrac{\der}{\der t} \templateflow{\vfield,\intenfield}{t}(x) = \intenfield\bigl(t, \diffeoflow{\vfield}{0,t}(x)\bigr) & \\[0.75em]
  \templateflow{\vfield,\intenfield}{0}(x) = \template(x) & \\[0.75em]
  \diffeoflow{\vfield}{0,t} \in G_V \text{ is given by \cref{eq.FlowDiffeo}} & 
\end{cases}
\quad\text{for $x \in \signaldomain$ and $t \in [0,T]$.}
\end{equation}
One can show that \cref{eq:MetaFlow} has a unique solution $t \mapsto (\diffeoflow{\vfield}{0,t},\templateflow{\vfield,\intenfield}{t}) \in G_V \times \RecSpace$ \cite{Trouve:2005aa,Charon:2016aa}, so the above construction can be used for deforming images.

\subsection{Deformable templates by metamorphosis}
The aim here is to solve \cref{eq:InvProbDeforTemp} with time discretised data. 
Following \cite{Gris:2019aa}, the idea is to adopt the \emph{independent trajectory} approach outlined in \cref{sec:IndRegSpatioTempTimeDiscData}, so the inverse problem can be reformulated as a sequence of indirect registration problems \cref{eq:UnMatchedTrajectory}.
Hence, the task reduces to recovering and matching a template $\template$ independently to data $\data_j$ in the sense of joint reconstruction and registration (indirect registration).
One could here consider various approaches for indirect registration, see \cite{Yang:2013aa,Chen:2018aa} for surveys, and \cite{Gris:2019aa} uses metamorphosis for this step.

The above considerations lead to the following variational formulation: 
\begin{equation}\label{eq:MetaSpatioTemp}
  (\deforparamopt_1,\ldots,\deforparamopt_n) 
  \in \!\!\!\!\argmin_{\deforparam_1,\ldots, \deforparam_n \in V \times \RecSpace}
  \biggl\{
  \sum_{i=1}^{n} 
     \DataDiscrep\Bigl(\ForwardOp_j\bigl(\DeforOp_{\deforparam_j}(\template) \bigr),\data_i \Bigr)
   + \lambda \Vert \vfield \Vert_2^2 + \tau \Vert \intenfield \Vert_2^2
  \biggr\}.
\end{equation}
The template $\template \in \RecSpace$ and data $\data_1, \dots, \data_{n} \in \DataSpace$ are related to each other as in \cref{eq:InvProbTimeDiscr} and the deformation operator $\DeforOp_{\deforparam_j} \colon \RecSpace \to \RecSpace$, which is parametrised by $\deforparam_j := (\vfield(t_j,\Cdot),\intenfield(t_j,\Cdot)) \in V \times \RecSpace$, is given by the metamorphosis framework as 
\begin{equation}\label{eq:DeforOpMeta}
 \DeforOp_{\deforparam_j}(\template) := \diffeoflow{\vfield}{0,t_i}.\templateflow{\vfield,\intenfield}{t_i} 
   \quad\text{where $(\diffeoflow{\vfield}{0,t},\templateflow{\vfield,\intenfield}{t}) \in G_V \times \RecSpace$ solves \cref{eq:MetaFlow}.}
\end{equation}
The group action in \cref{eq:DeforOpMeta} is usually the geometric one in \cref{eq:GeometricGroupAction}.

The approach taken in \cite{Gris:2019aa} is based on solving \cref{eq:MetaSpatioTemp} by a scheme that intertwines updates of the image with updates of the deformation parameter.
The latter involves solving an indirect registration problem and a key part of \cite{Gris:2019aa} is to show that indirect registration by metamorphosis has a solution \cite[Proposition~4]{Gris:2019aa} (existence) that is continuous w.r.t. data \cite[Proposition~5]{Gris:2019aa} (stability) and convergent \cite[Proposition~6]{Gris:2019aa}.
As such, the updates of the deformation parameter by metamorphosis based indirect registration is a well defined regularisation method in the sense of \cite{grasmair2010generalized}.
Likewise, the updates of the image is by a variational method that defines a well defined regularisation method, so both updates of the intertwined scheme for solving \cref{eq:MetaSpatioTemp} are by regularisation methods. 

\begin{figure}[ht]
   \begin{minipage}[t]{0.2\textwidth}%
     \centering
     \includegraphics[trim=45 20 5 5, clip, width=\textwidth]{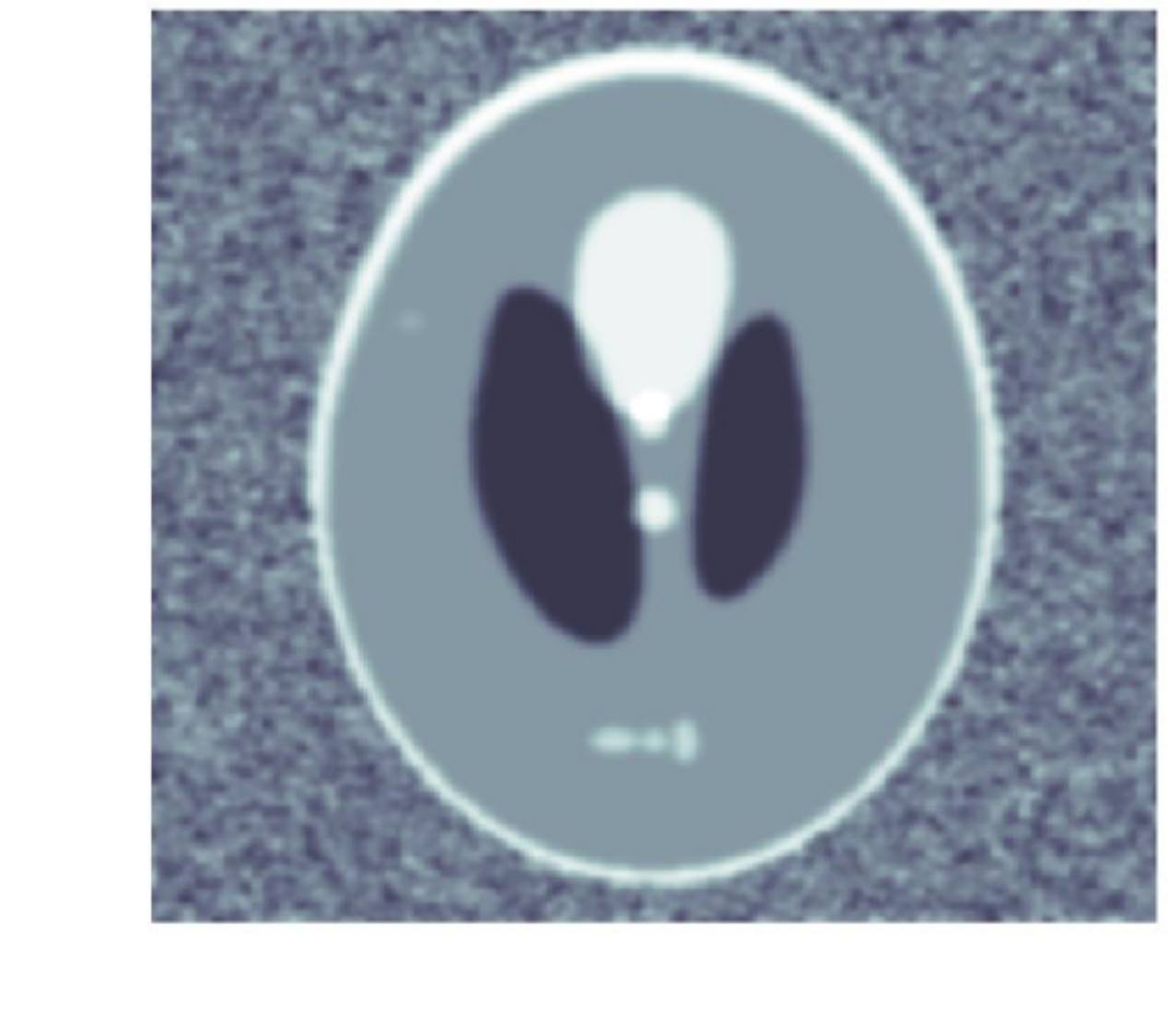}
   \end{minipage}%
   \begin{minipage}[t]{0.2\textwidth}%
     \centering
     \includegraphics[trim=45 20 5 5, clip, width=\textwidth]{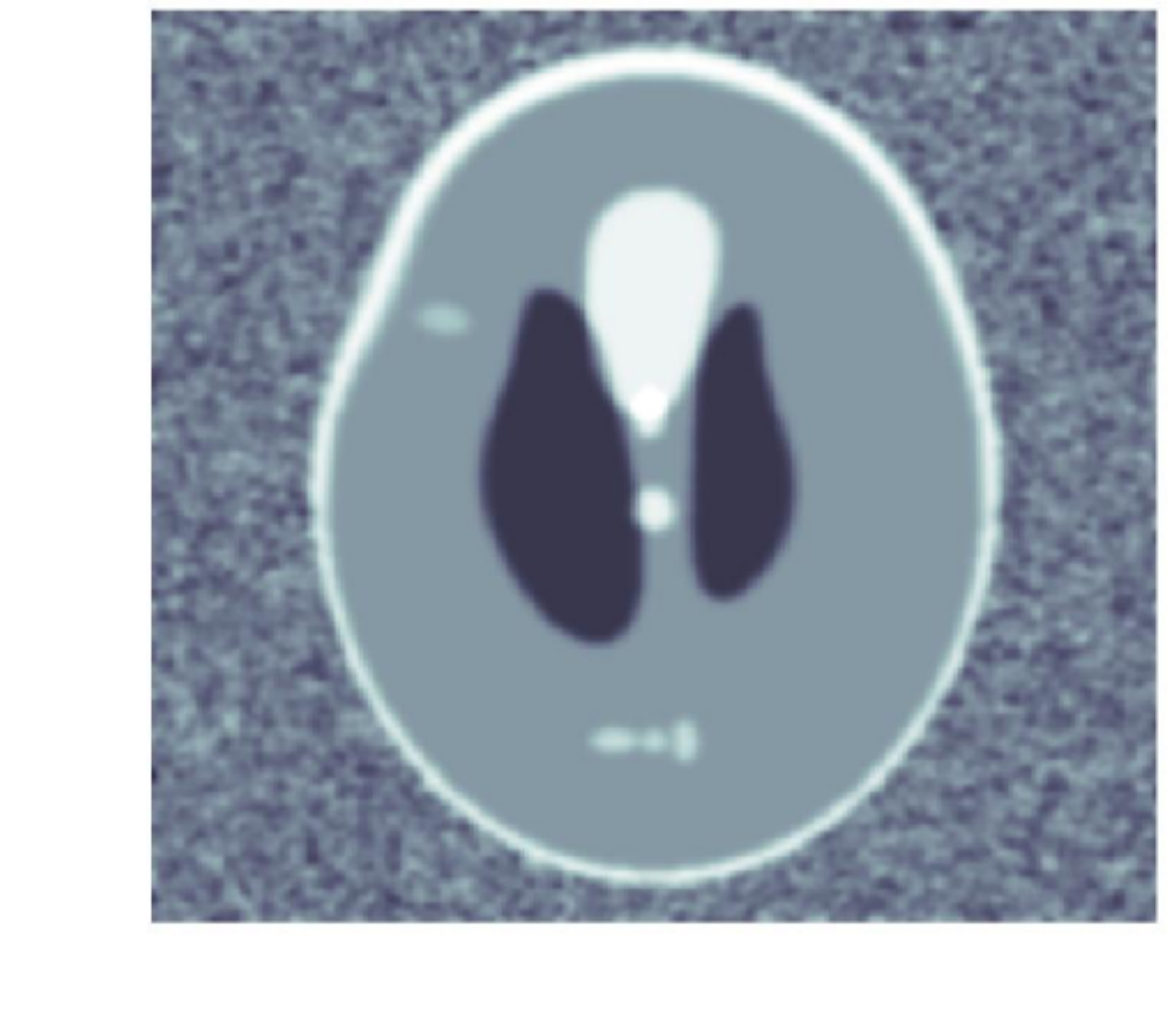}
   \end{minipage}%
   \begin{minipage}[t]{0.2\textwidth}%
     \centering
     \includegraphics[trim=45 20 5 5, clip, width=\textwidth]{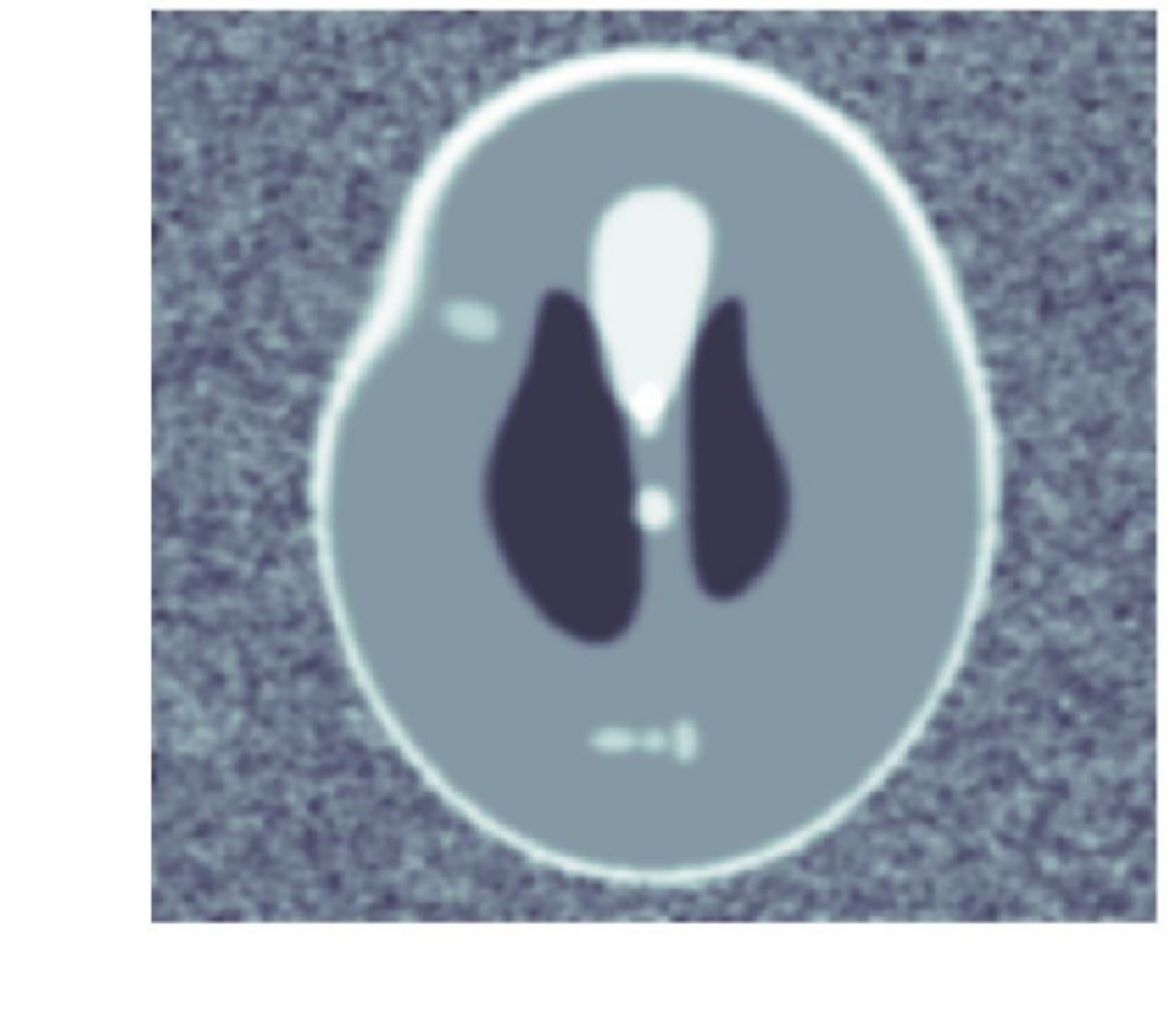}
   \end{minipage}%
   \begin{minipage}[t]{0.2\textwidth}%
     \centering
     \includegraphics[trim=45 20 5 5, clip, width=\textwidth]{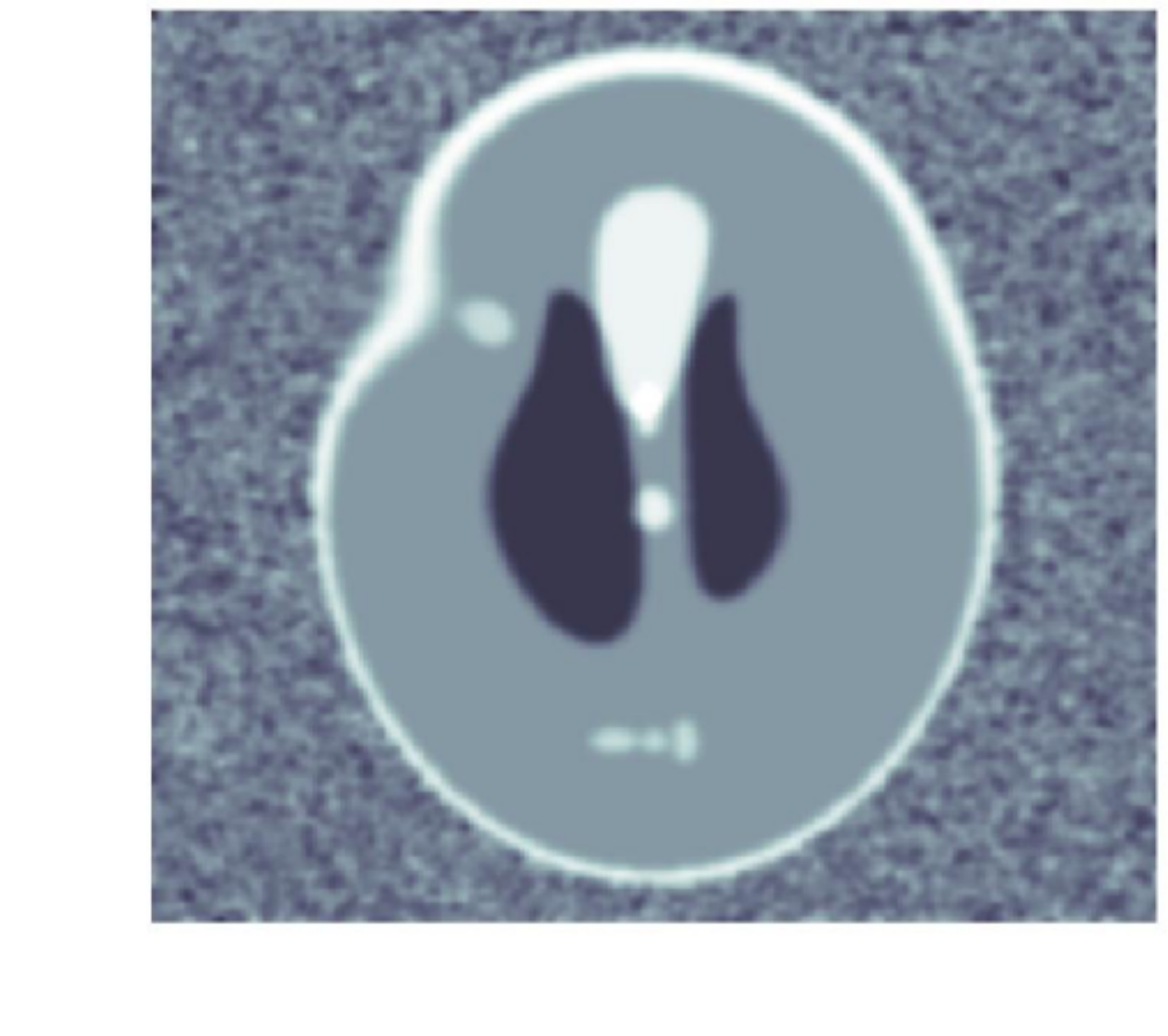}
   \end{minipage}%
   \begin{minipage}[t]{0.2\textwidth}%
     \centering
     \includegraphics[trim=45 20 5 5, clip, width=\textwidth]{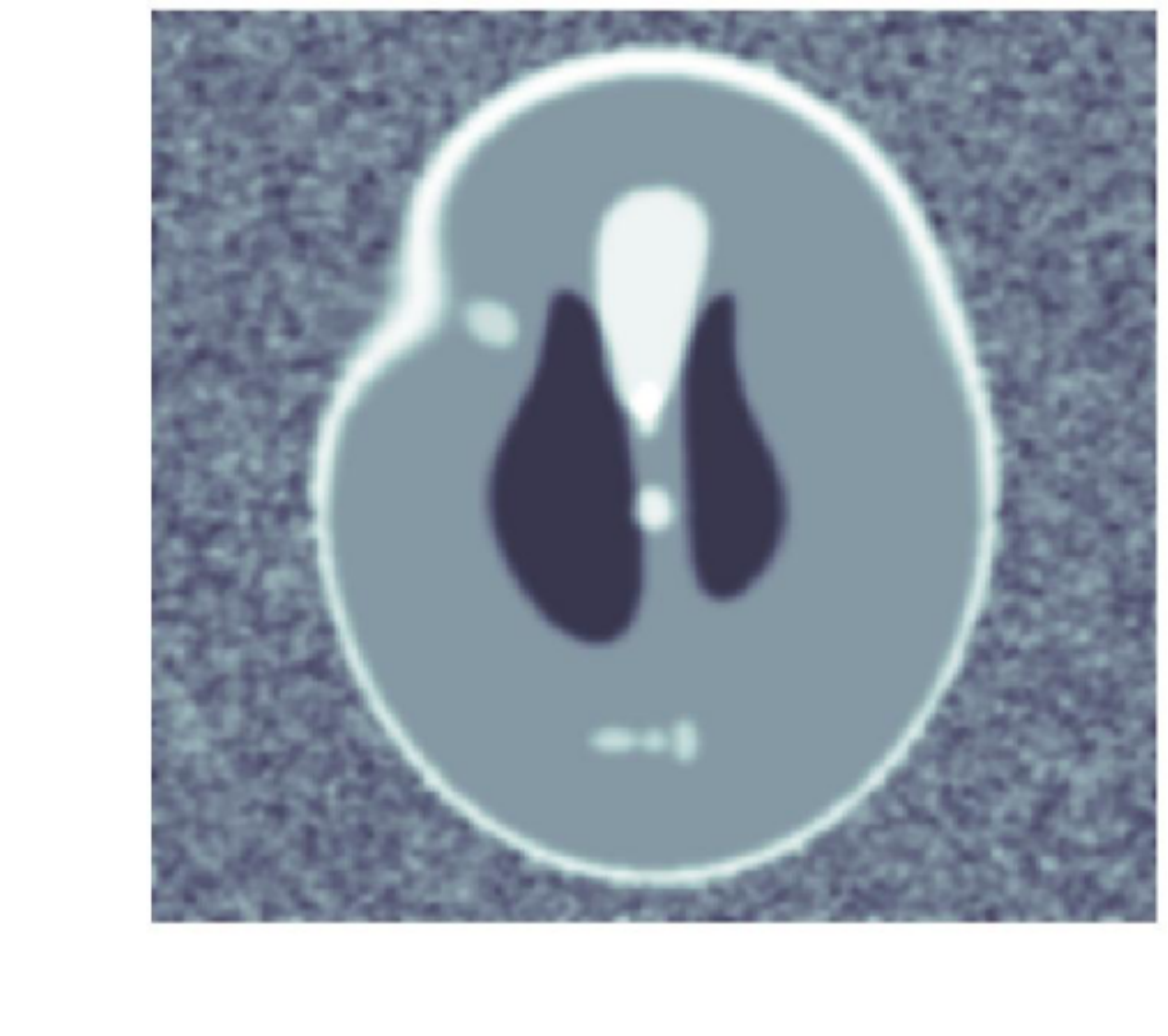}
   \end{minipage}%
   \vskip-0.25\baselineskip
   Ground truth (unknown) $256 \times 256$ pixel grey scale spatiotemporal target image.
   \\[0.5em]
   \begin{minipage}[t]{0.2\textwidth}%
     \centering
     \includegraphics[trim=45 20 5 5, clip, width=\textwidth]{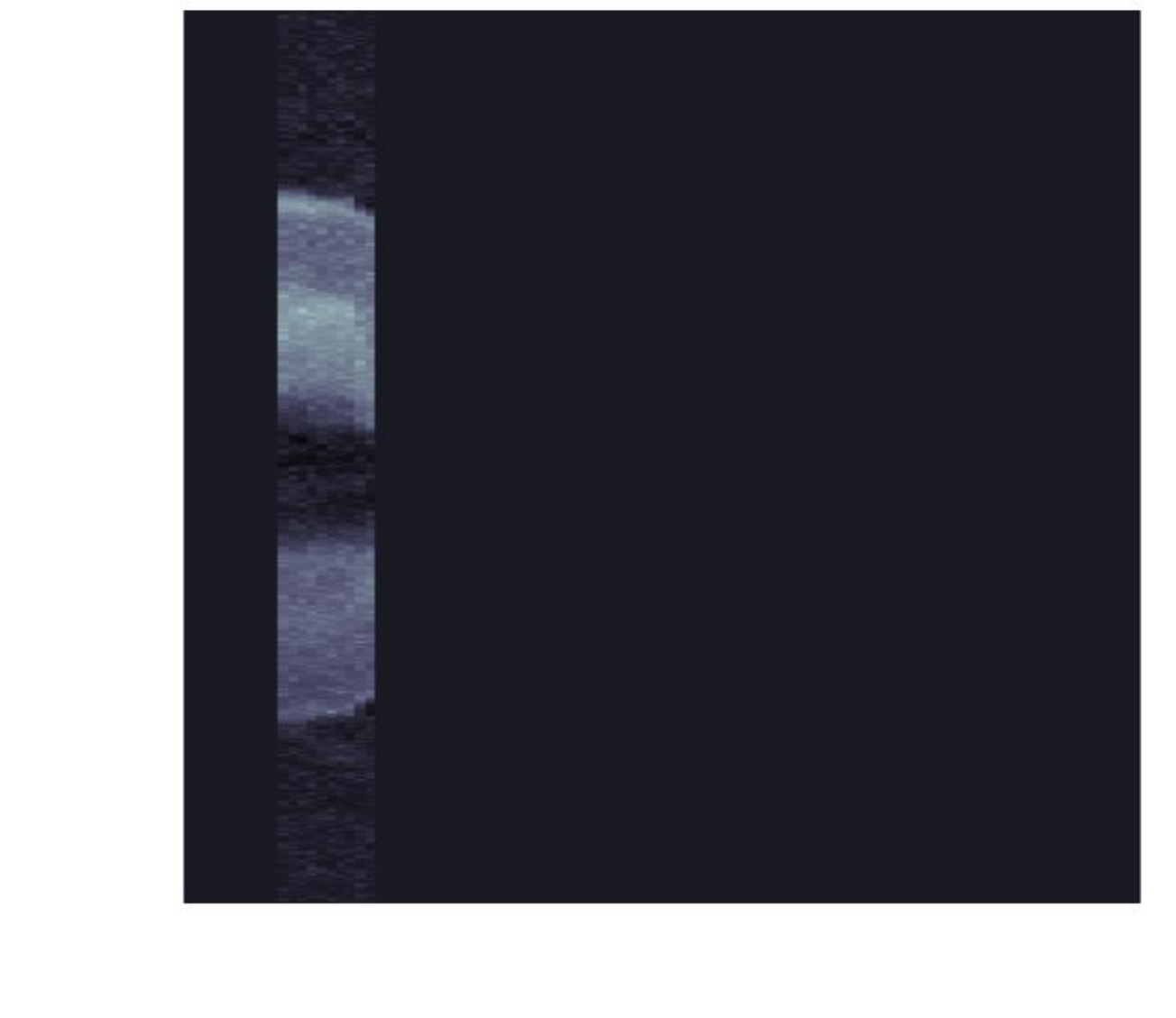}
   \end{minipage}%
   \begin{minipage}[t]{0.2\textwidth}%
     \centering
     \includegraphics[trim=45 20 5 5, clip, width=\textwidth]{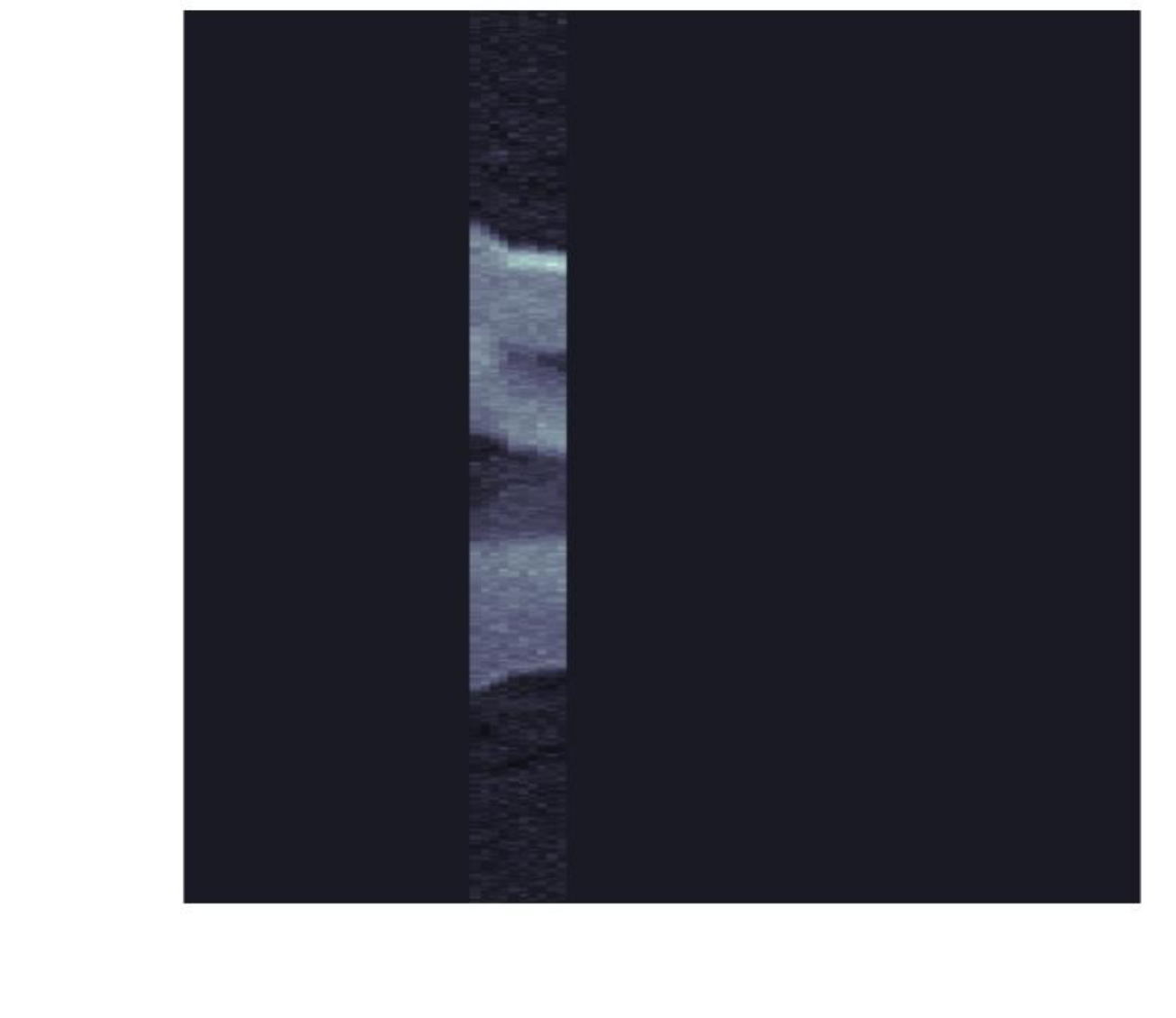}
   \end{minipage}%
   \begin{minipage}[t]{0.2\textwidth}%
     \centering
     \includegraphics[trim=45 20 5 5, clip, width=\textwidth]{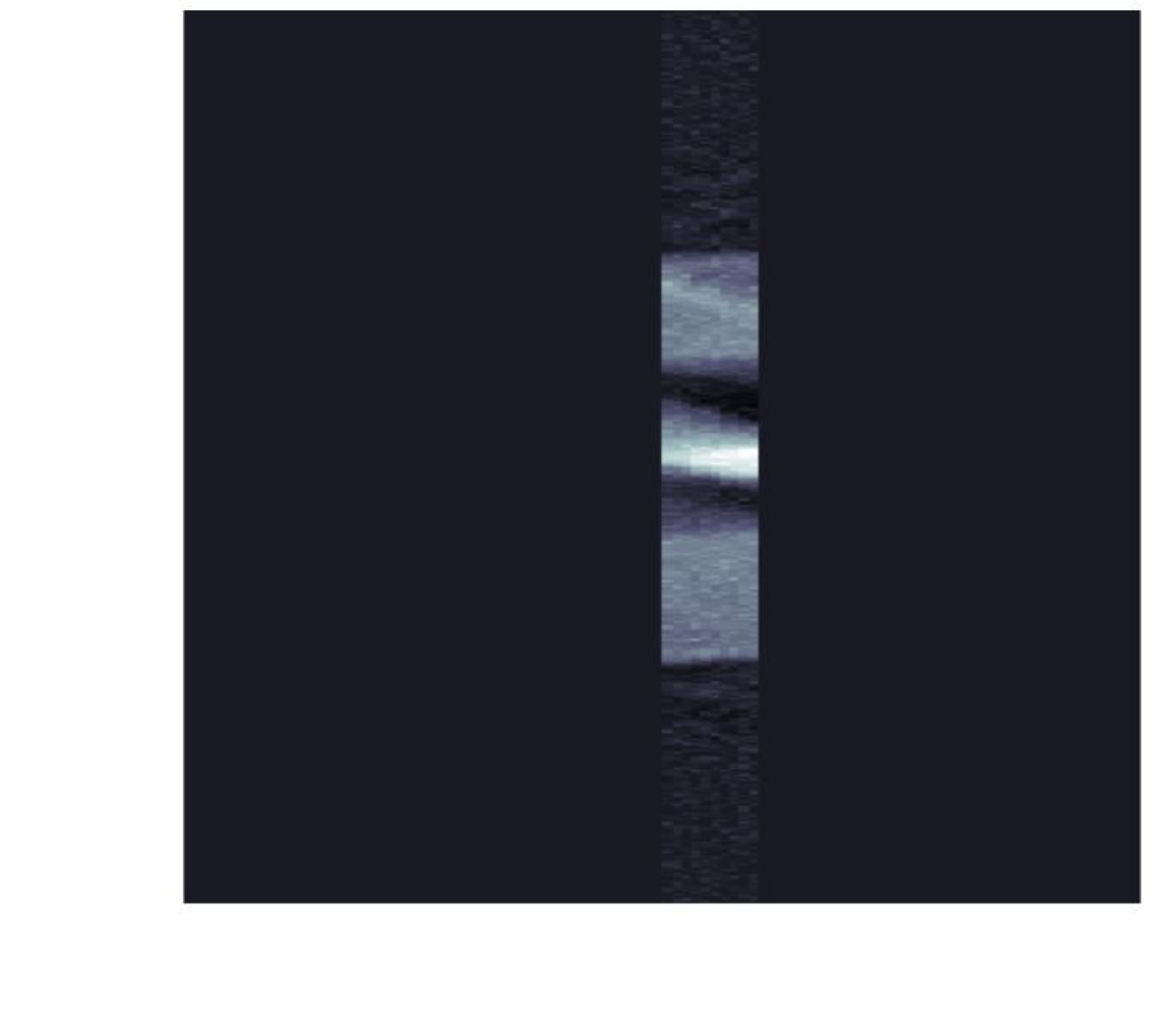}
   \end{minipage}%
   \begin{minipage}[t]{0.2\textwidth}%
     \centering
     \includegraphics[trim=45 20 5 5, clip, width=\textwidth]{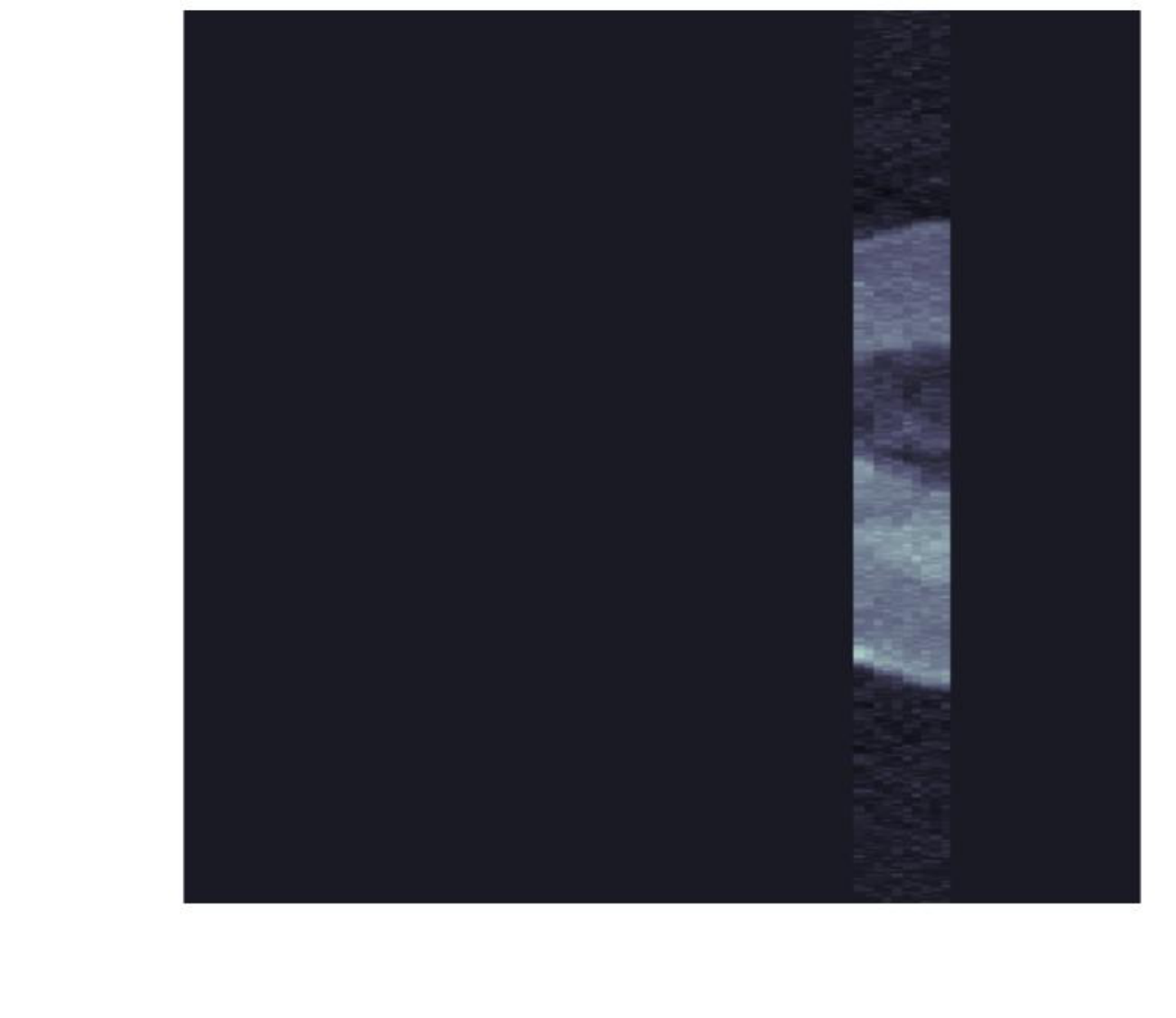}
   \end{minipage}%
   \begin{minipage}[t]{0.2\textwidth}%
     \centering
     \includegraphics[trim=45 20 5 5, clip, width=\textwidth]{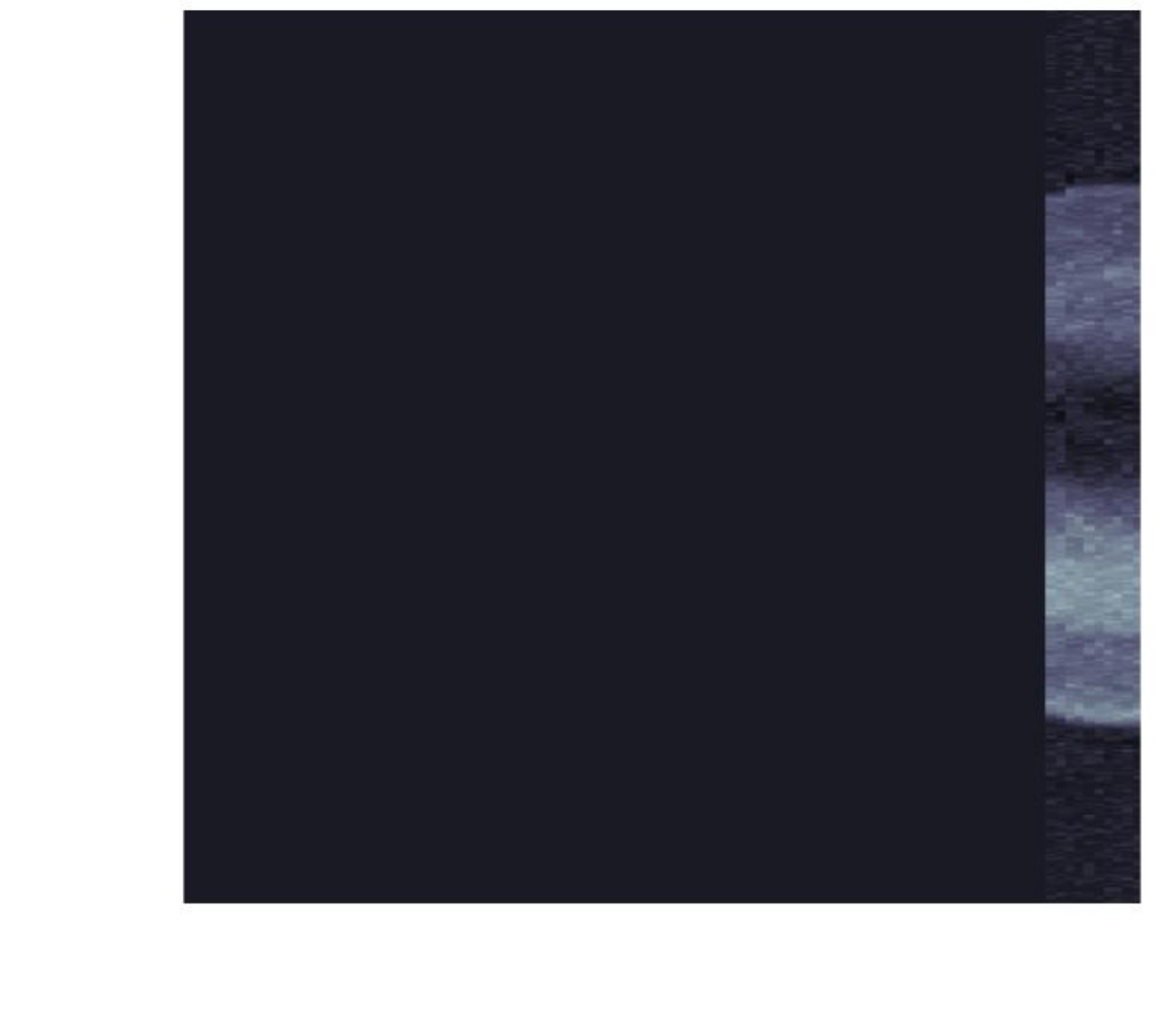}
   \end{minipage}%
   \vskip-0.25\baselineskip
   Gated noisy tomographic projection data of spatiotemporal target image.
   We sample the parallel beam ray transform at time $t_i$ using 10~angles randomly distributed in $[(i-1) \pi / 10, i \pi / 10]$. Data is corrupted with Poisson noise.
   \\[0.75em]
   \begin{minipage}[t]{0.2\textwidth}%
     \centering
     \includegraphics[trim=45 20 5 5, clip, width=\textwidth]{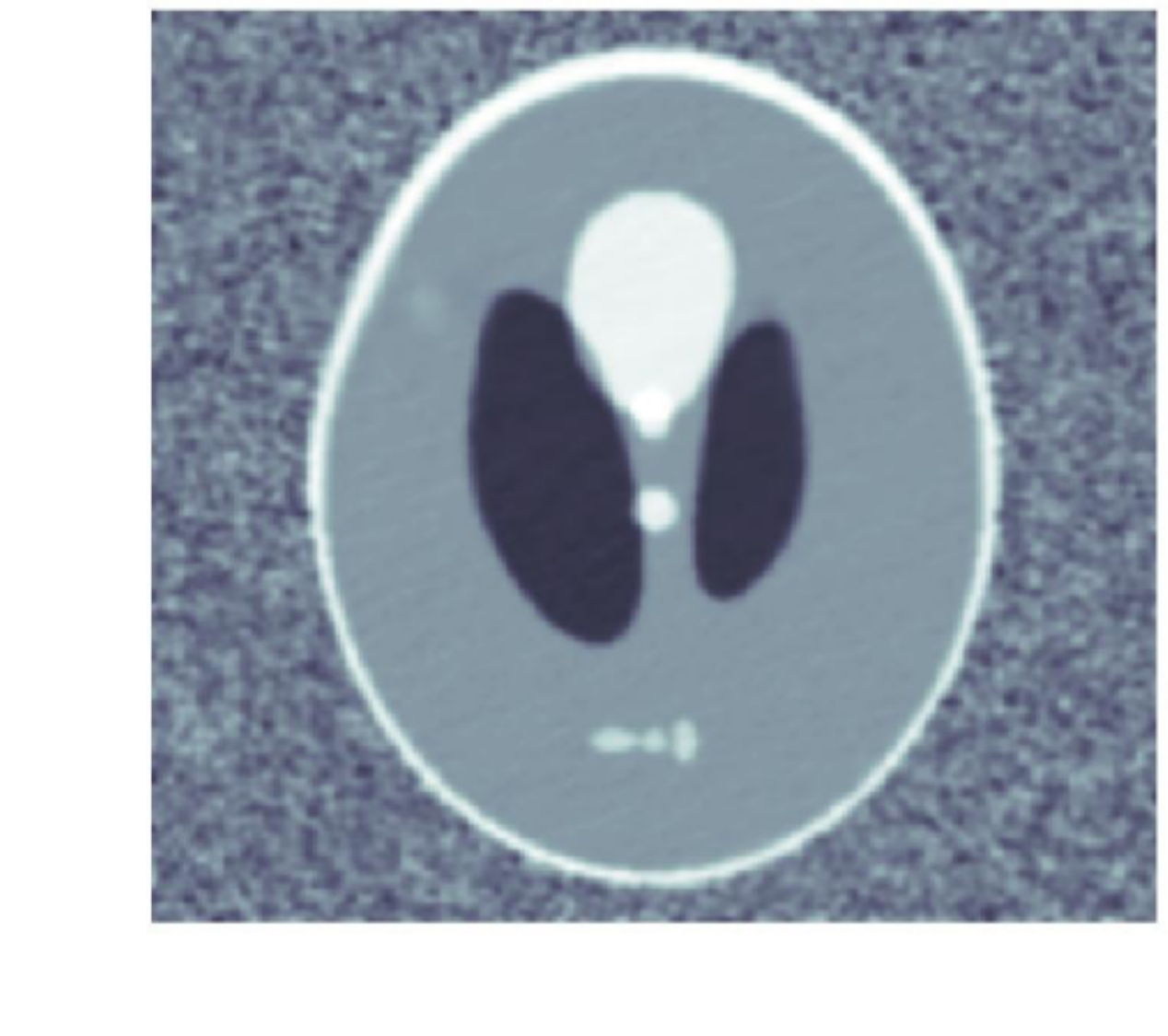}
   \end{minipage}%
   \begin{minipage}[t]{0.2\textwidth}%
     \centering
     \includegraphics[trim=45 20 5 5, clip, width=\textwidth]{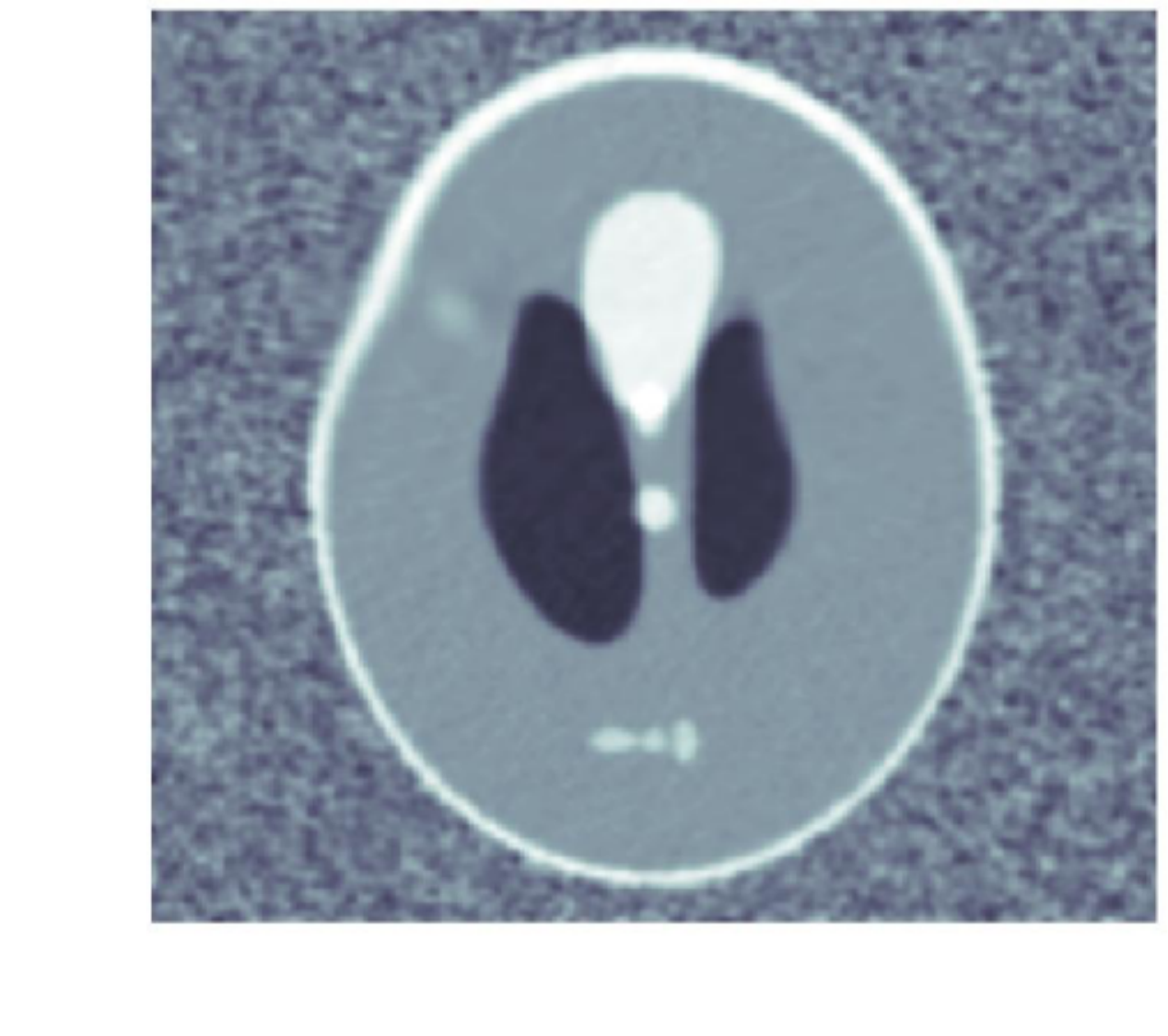}
   \end{minipage}%
   \begin{minipage}[t]{0.2\textwidth}%
     \centering
     \includegraphics[trim=45 20 5 5, clip, width=\textwidth]{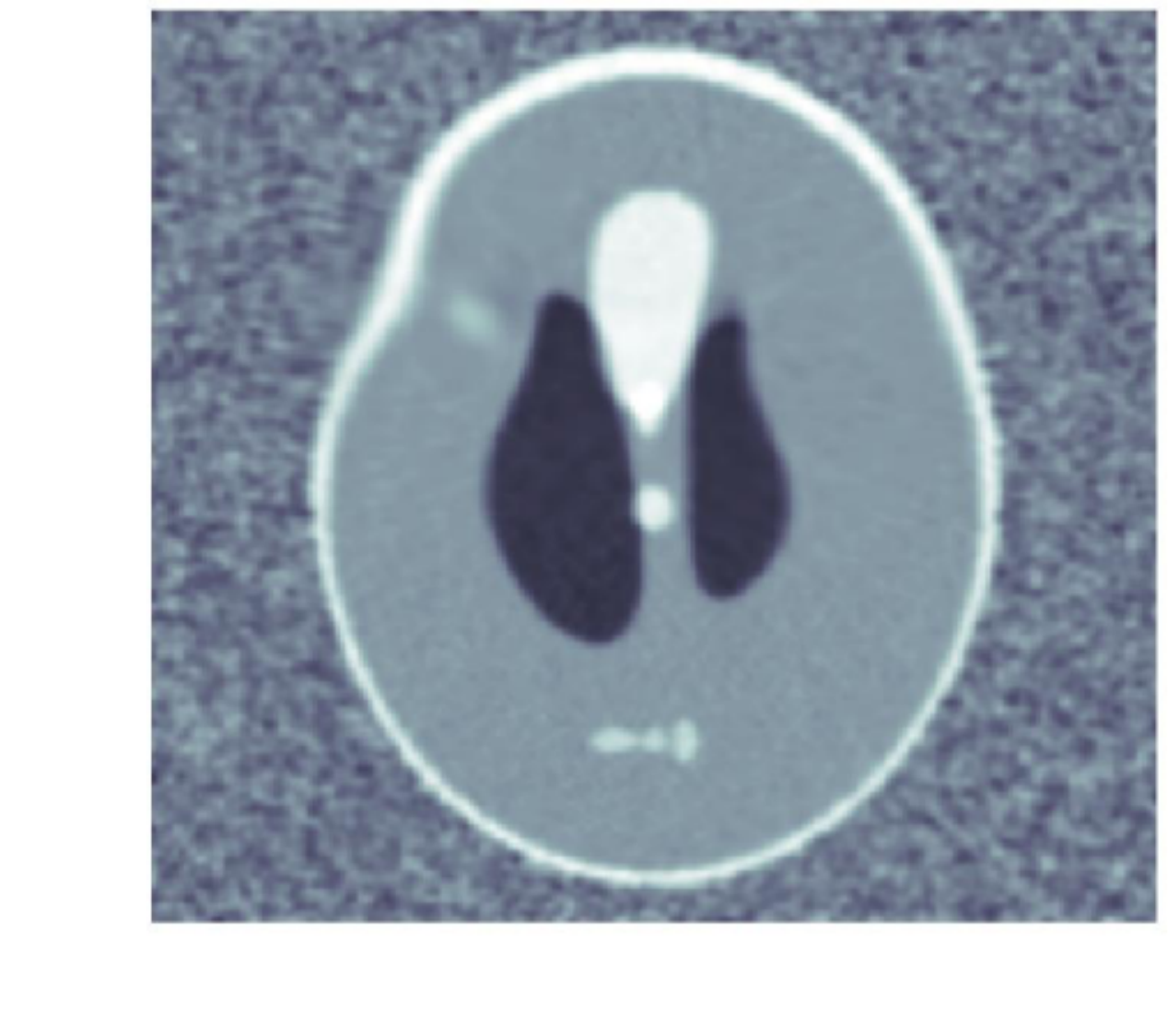}
   \end{minipage}%
   \begin{minipage}[t]{0.2\textwidth}%
     \centering
     \includegraphics[trim=45 20 5 5, clip, width=\textwidth]{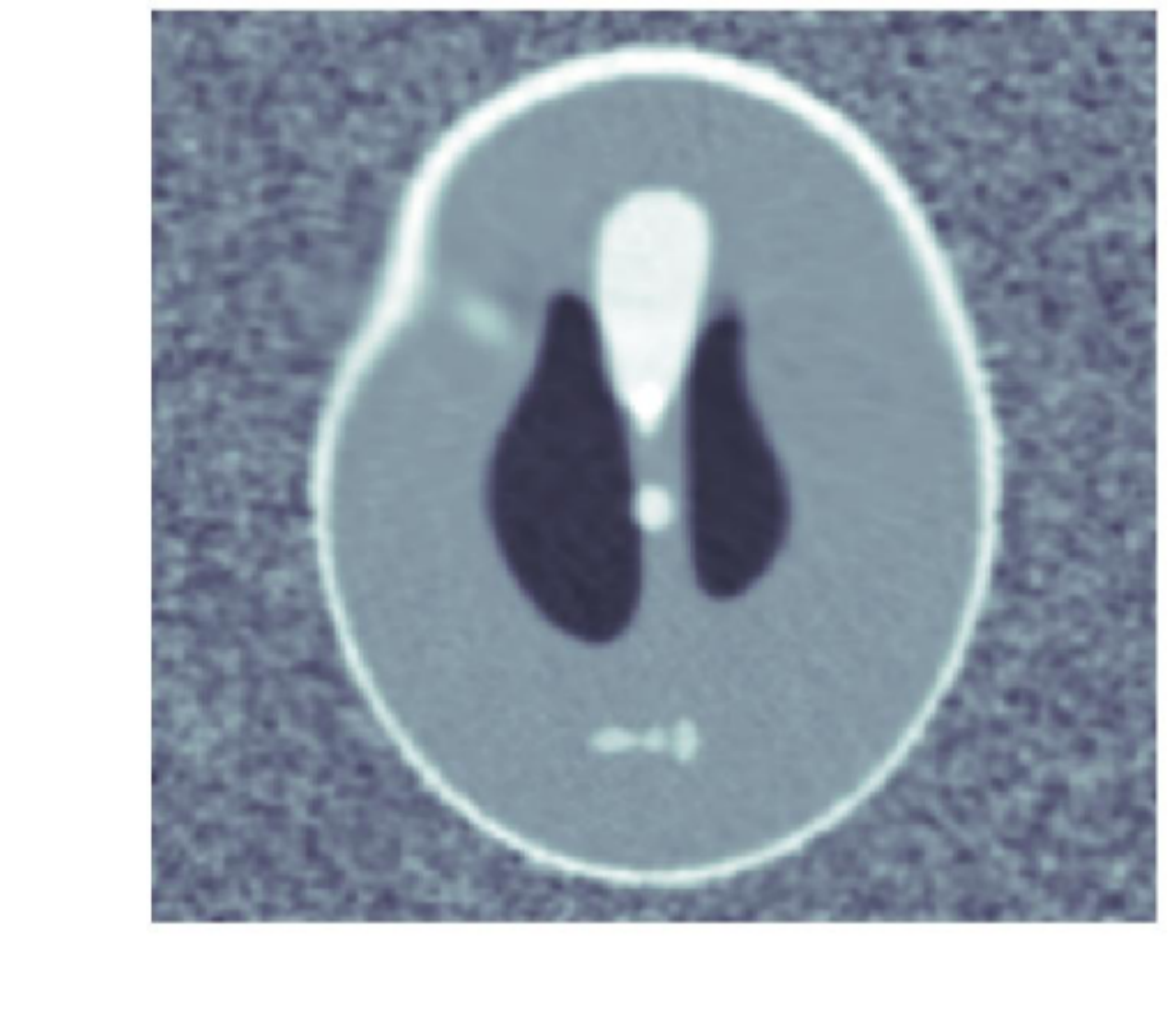}
   \end{minipage}%
   \begin{minipage}[t]{0.2\textwidth}%
     \centering
     \includegraphics[trim=45 20 5 5, clip, width=\textwidth]{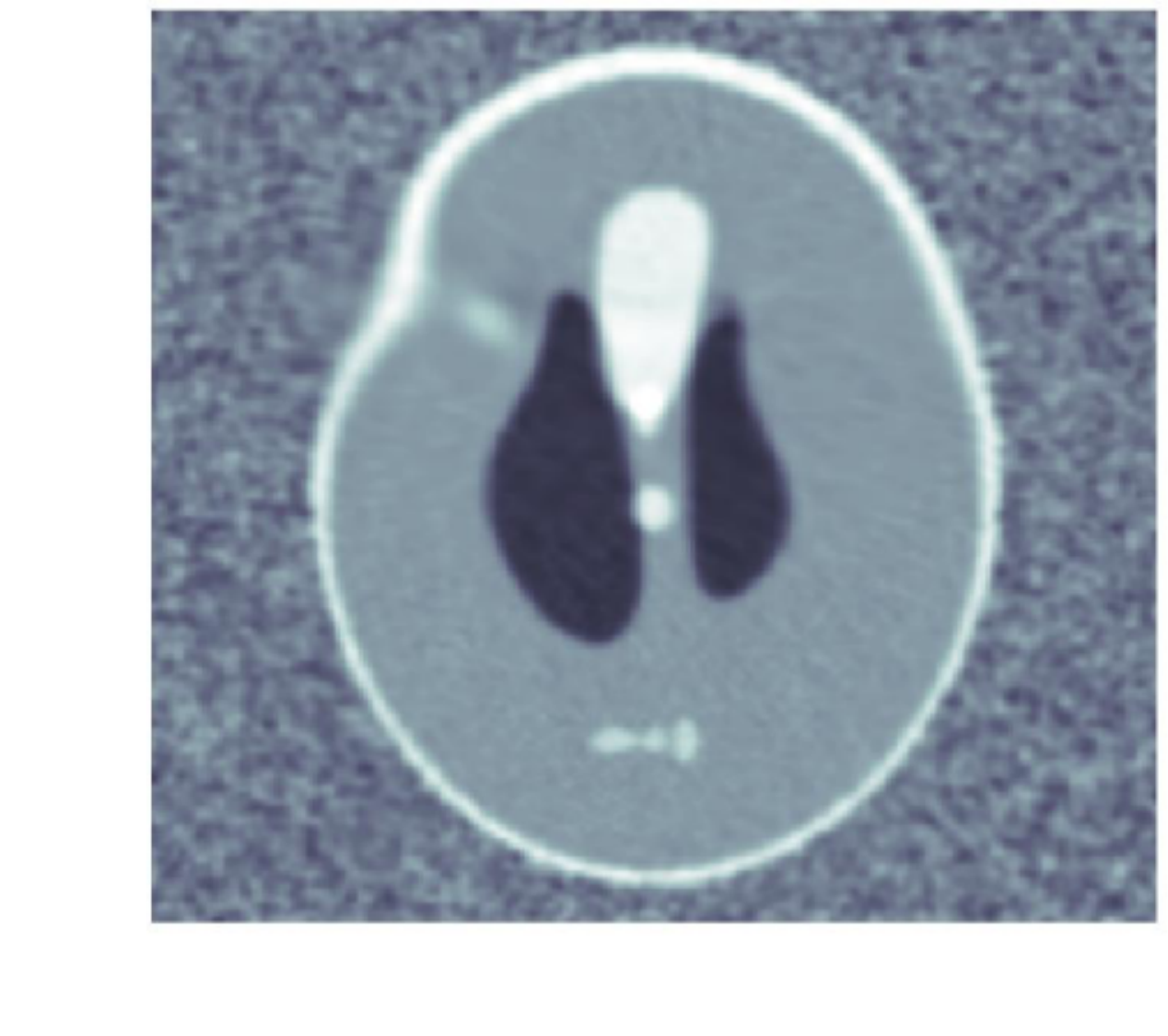}
   \end{minipage}%
   \vskip-0.25\baselineskip
   Image trajectory obtained by solving \cref{eq:MetaSpatioTemp} 
   \\[0.75em]
   \begin{minipage}[t]{0.2\textwidth}%
     \centering
     \includegraphics[trim=45 20 5 5, clip, width=\textwidth]{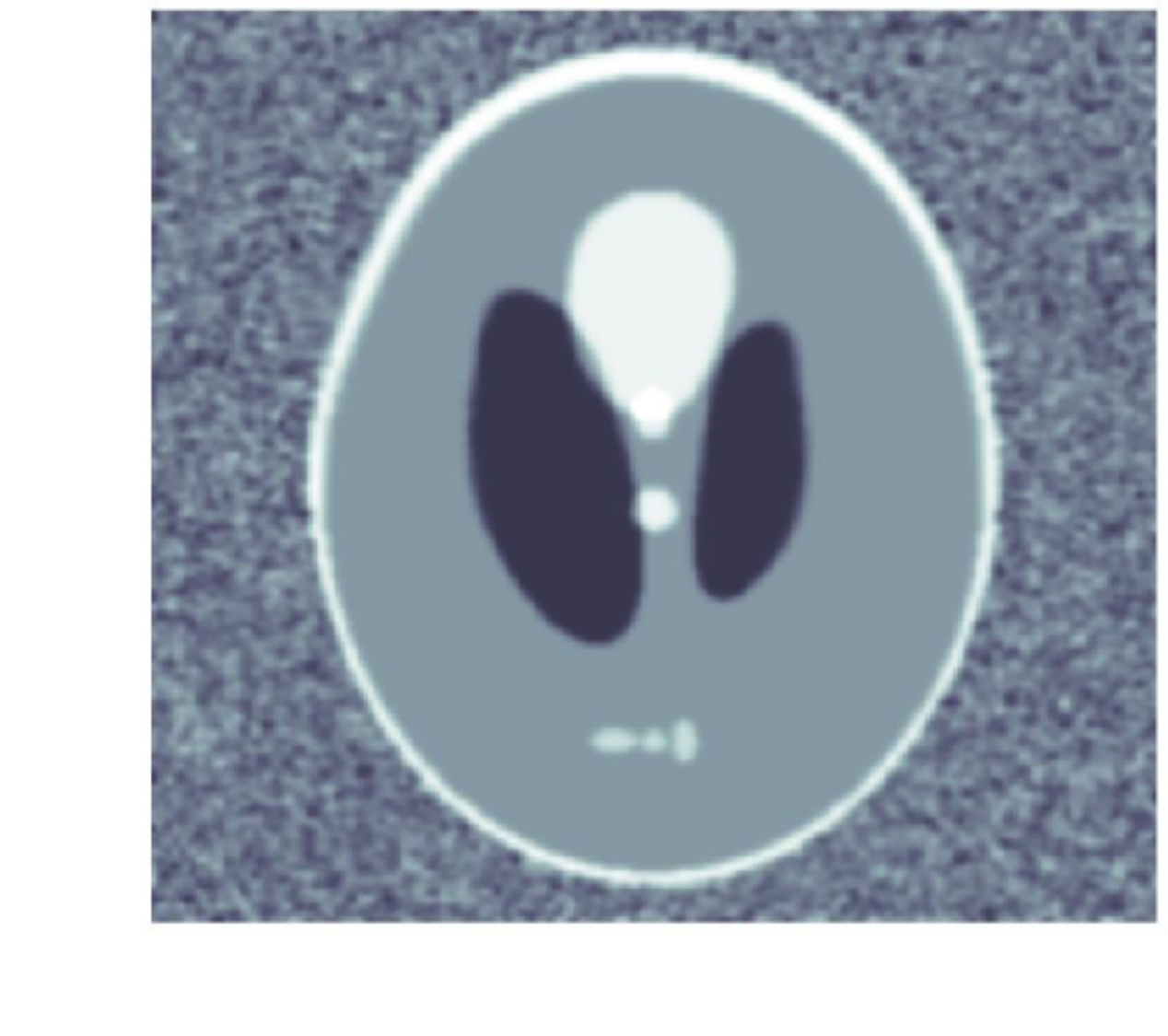}
   \end{minipage}%
   \begin{minipage}[t]{0.2\textwidth}%
     \centering
     \includegraphics[trim=45 20 5 5, clip, width=\textwidth]{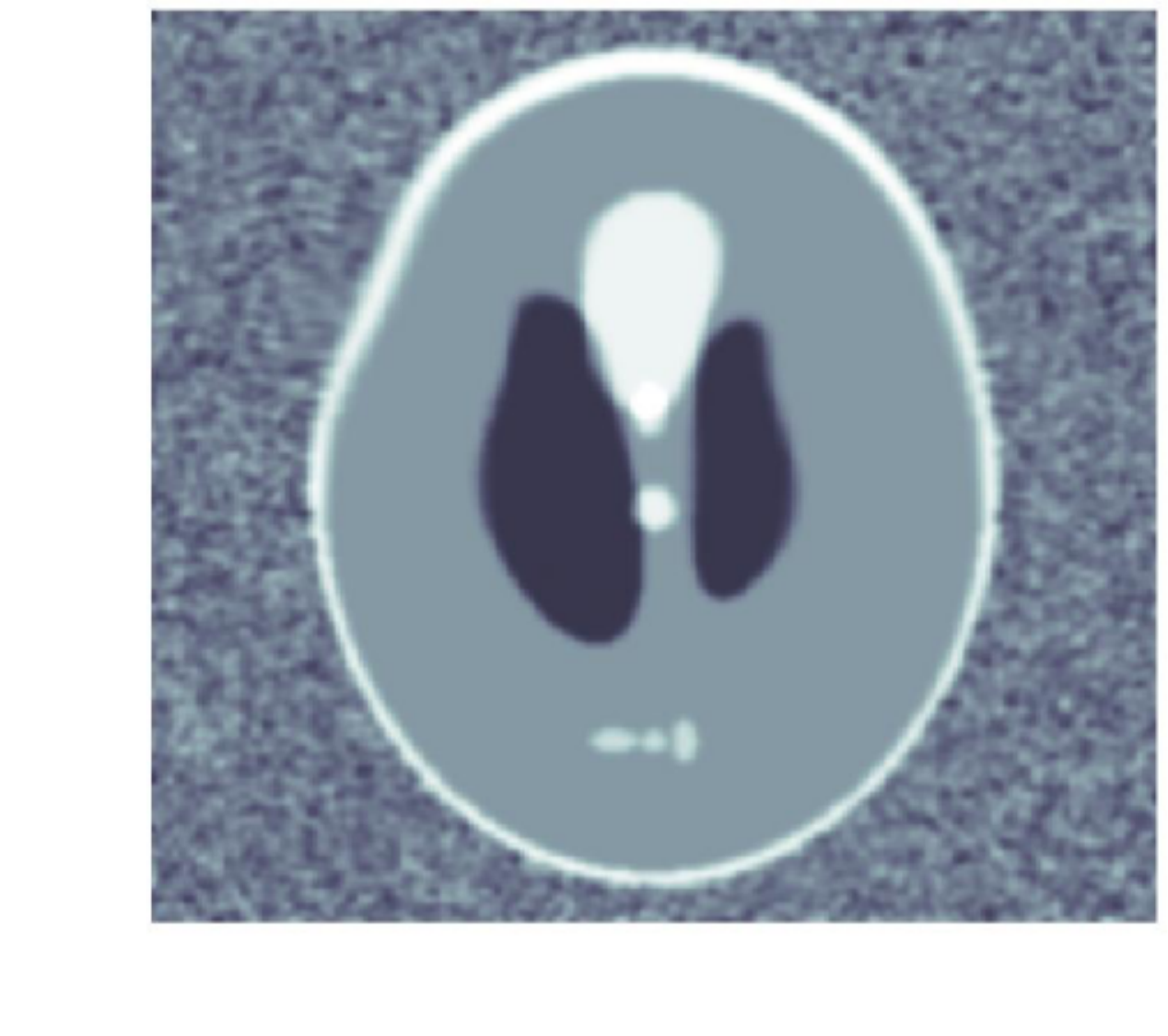}
   \end{minipage}%
   \begin{minipage}[t]{0.2\textwidth}%
     \centering
     \includegraphics[trim=45 20 5 5, clip, width=\textwidth]{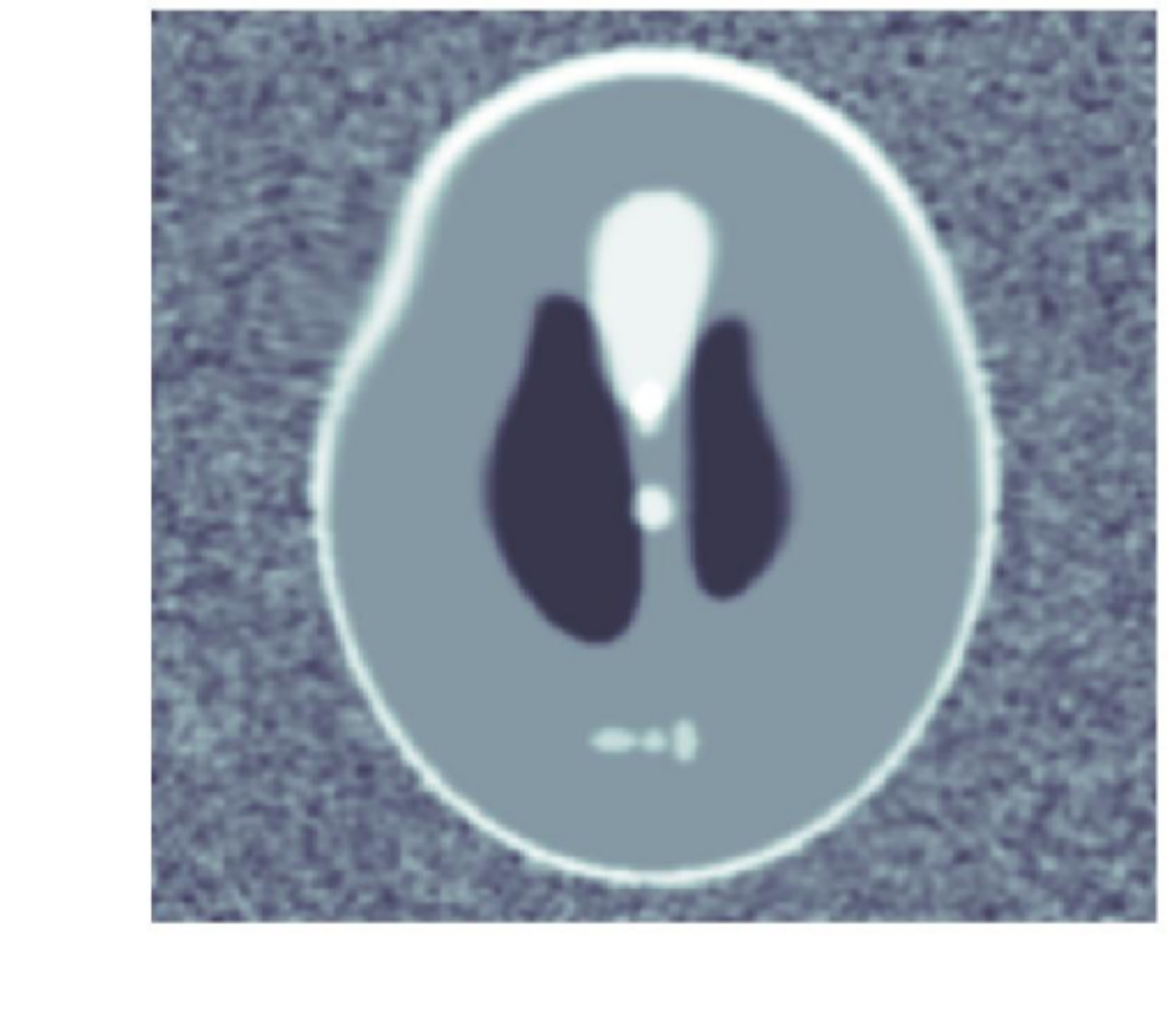}
   \end{minipage}%
   \begin{minipage}[t]{0.2\textwidth}%
     \centering
     \includegraphics[trim=45 20 5 5, clip, width=\textwidth]{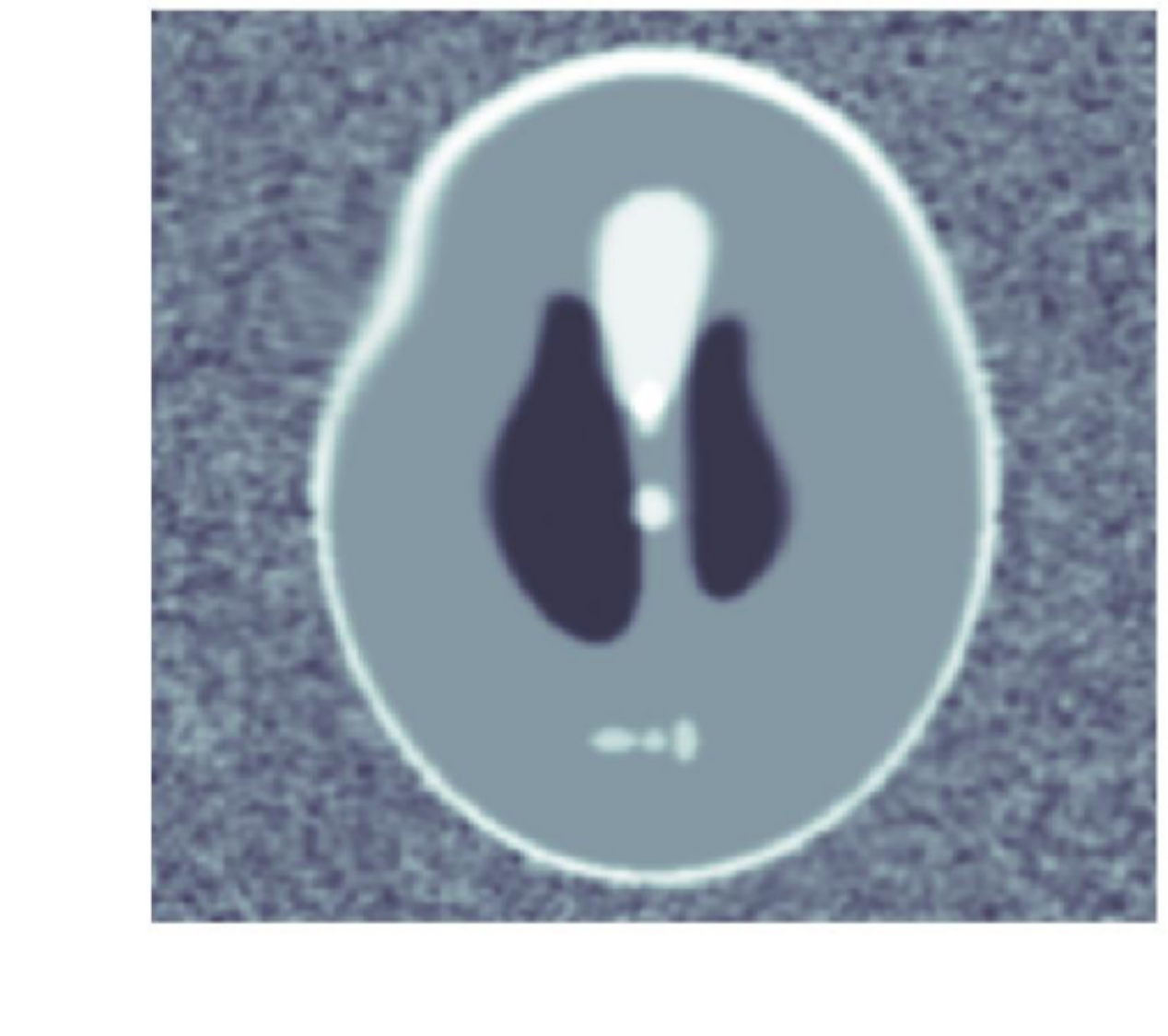}
   \end{minipage}%
   \begin{minipage}[t]{0.2\textwidth}%
     \centering
     \includegraphics[trim=45 20 5 5, clip, width=\textwidth]{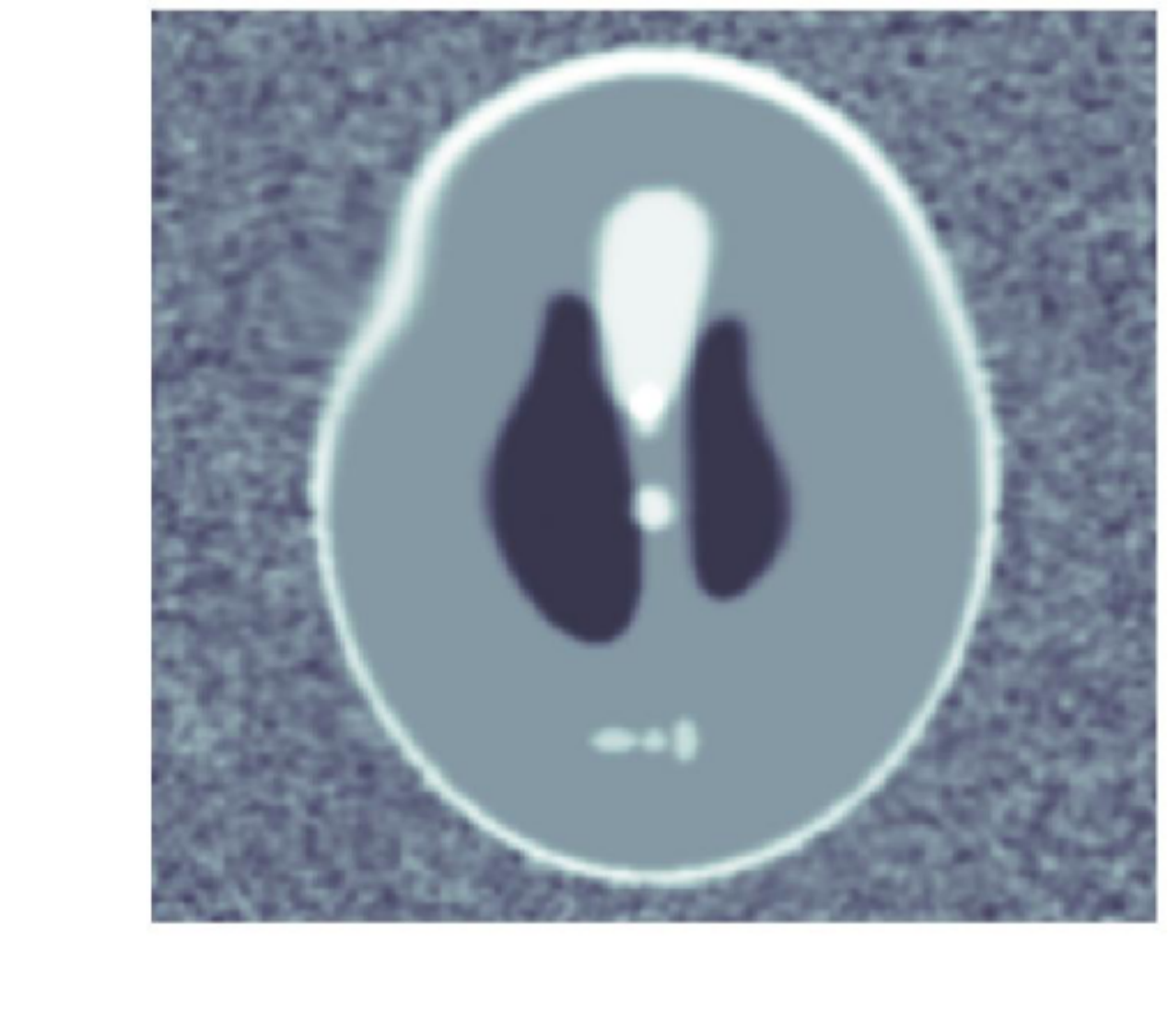}
   \end{minipage}%
   \vskip-0.25\baselineskip
   Shape trajectory obtained by solving \cref{eq:MetaSpatioTemp} 
   \\[0.75em]
   \begin{minipage}[t]{0.2\textwidth}%
     \centering
     \includegraphics[trim=45 20 5 5, clip, width=\textwidth]{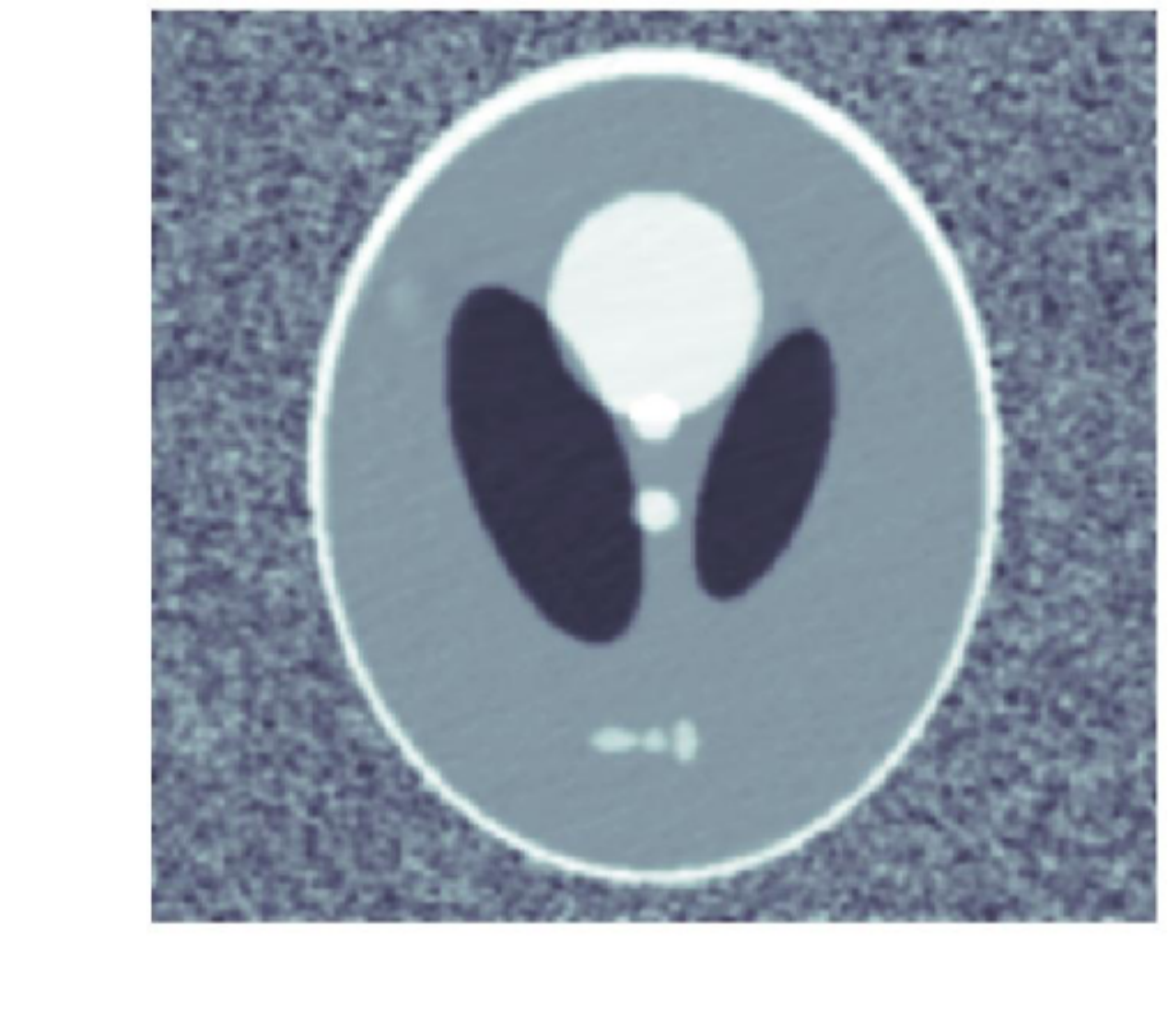}
   \end{minipage}%
   \begin{minipage}[t]{0.2\textwidth}%
     \centering
     \includegraphics[trim=45 20 5 5, clip, width=\textwidth]{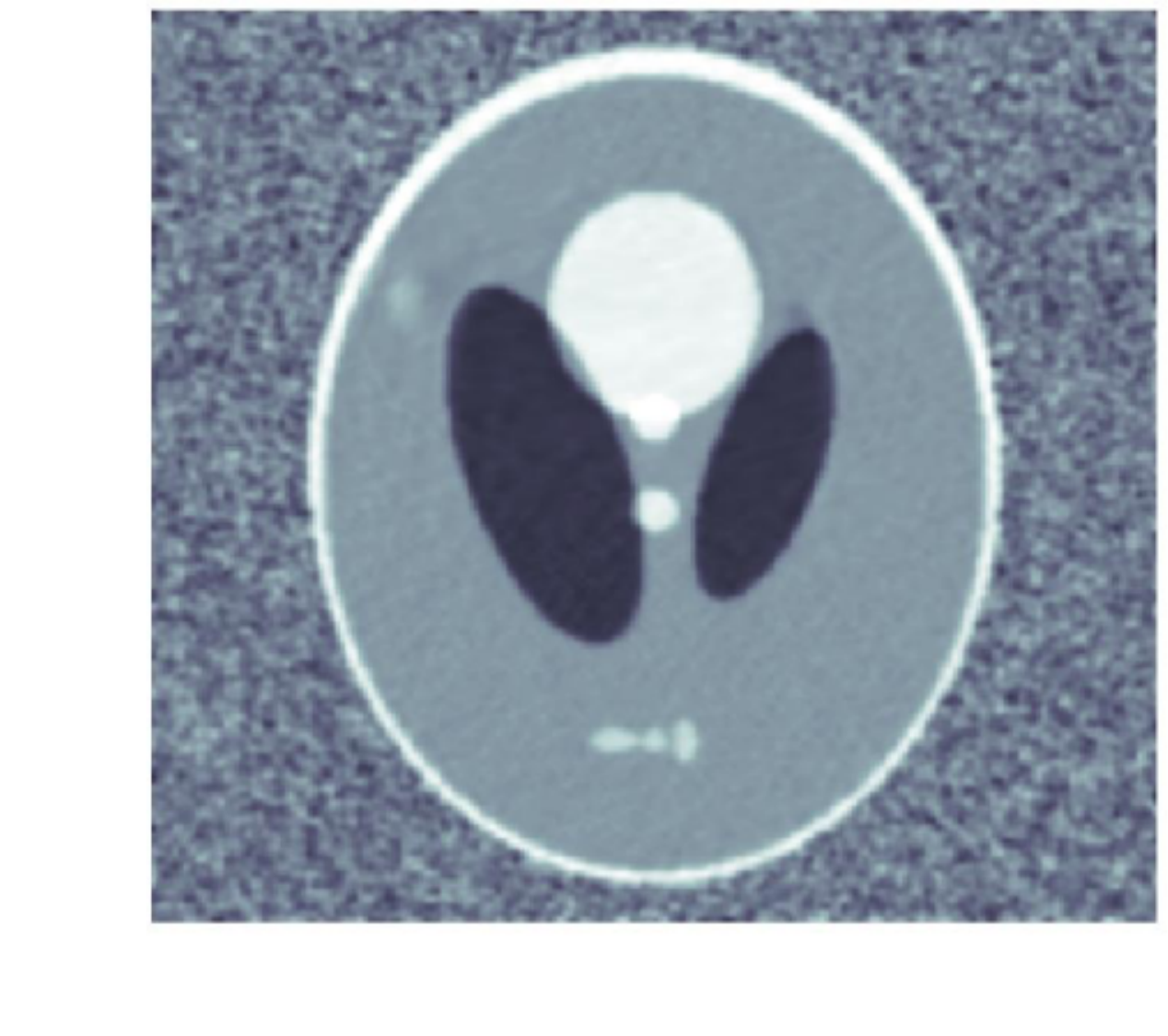}
   \end{minipage}%
   \begin{minipage}[t]{0.2\textwidth}%
     \centering
     \includegraphics[trim=45 20 5 5, clip, width=\textwidth]{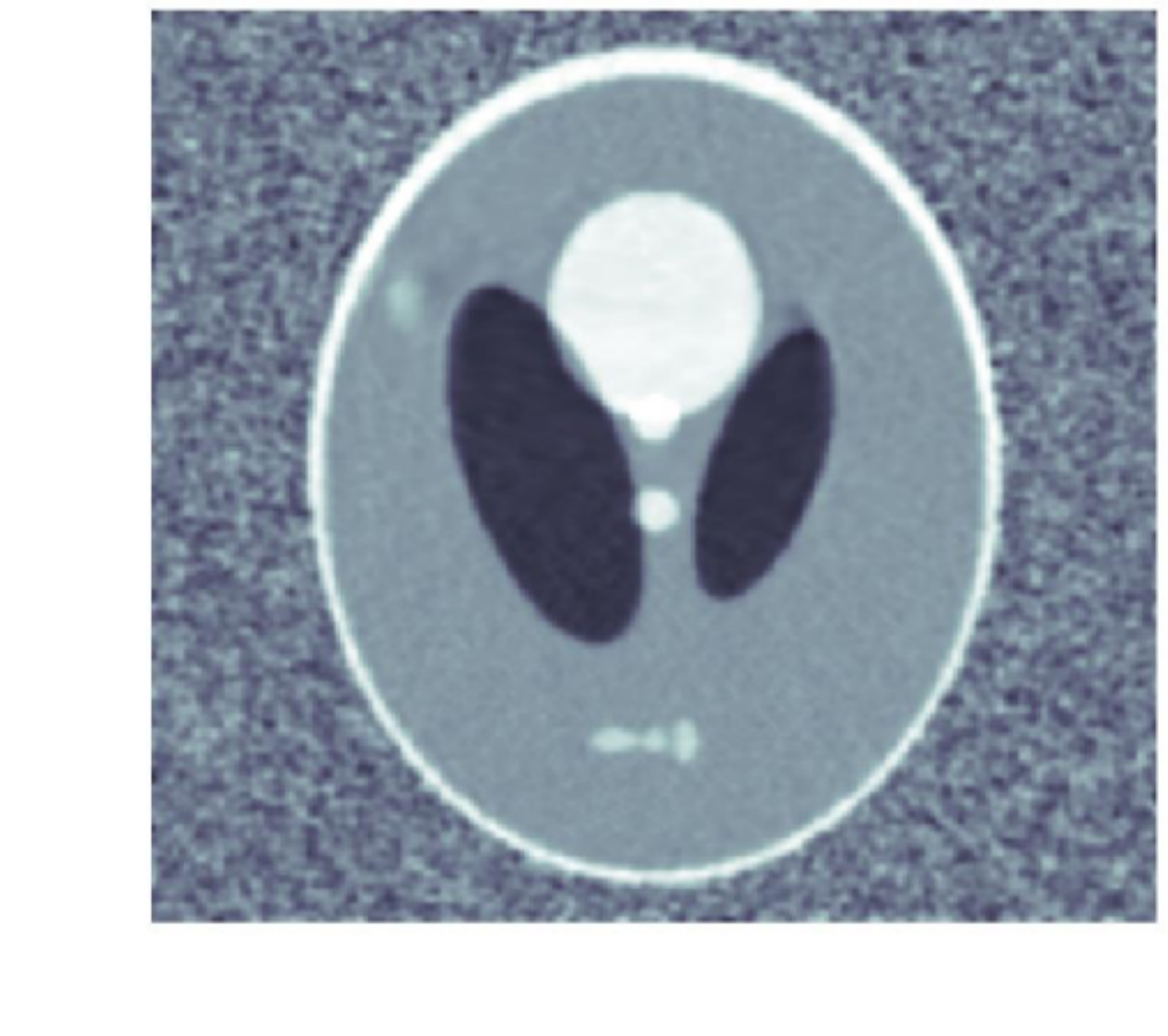}
   \end{minipage}%
   \begin{minipage}[t]{0.2\textwidth}%
     \centering
     \includegraphics[trim=45 20 5 5, clip, width=\textwidth]{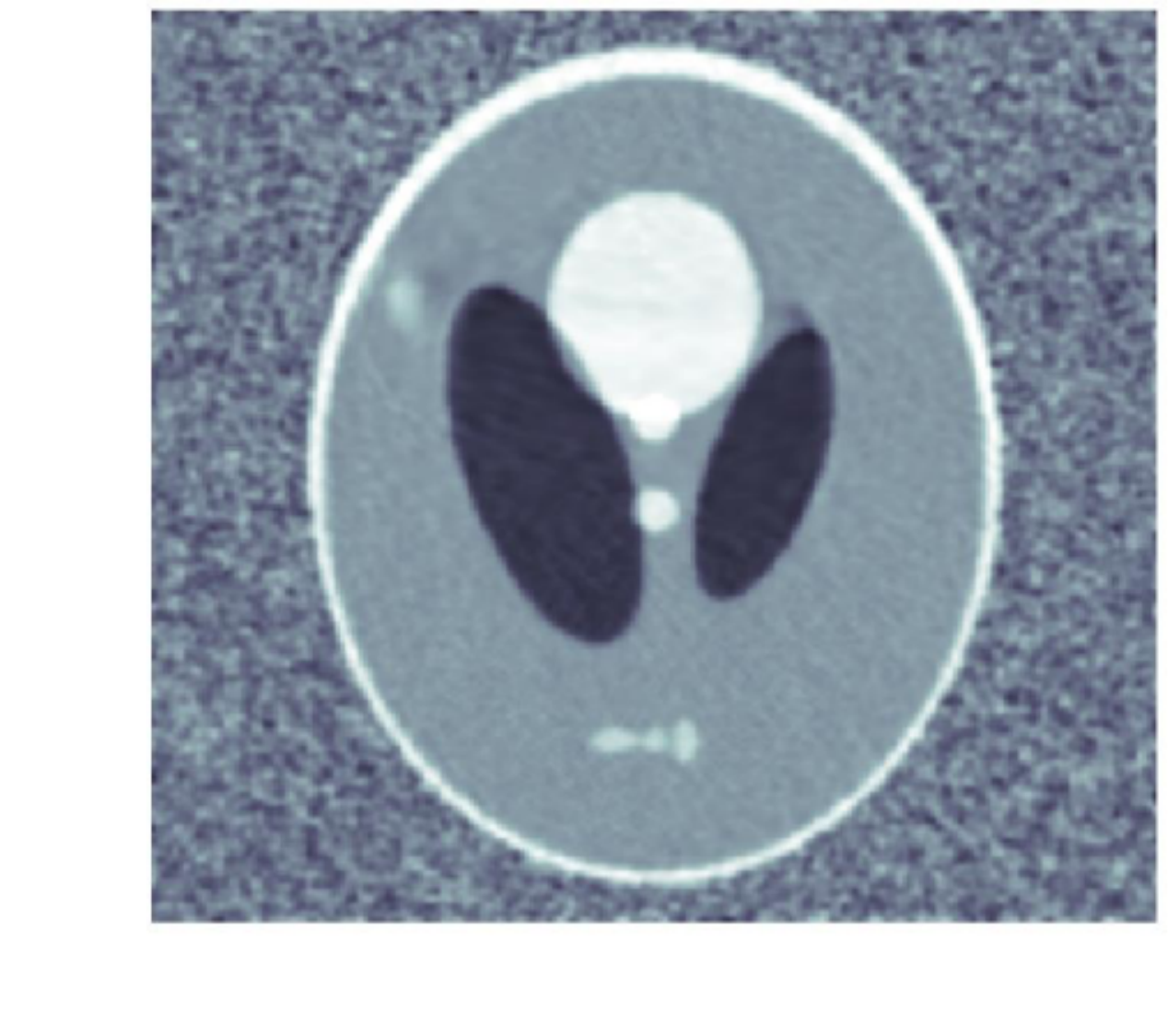}
   \end{minipage}%
   \begin{minipage}[t]{0.2\textwidth}%
     \centering
     \includegraphics[trim=45 20 5 5, clip, width=\textwidth]{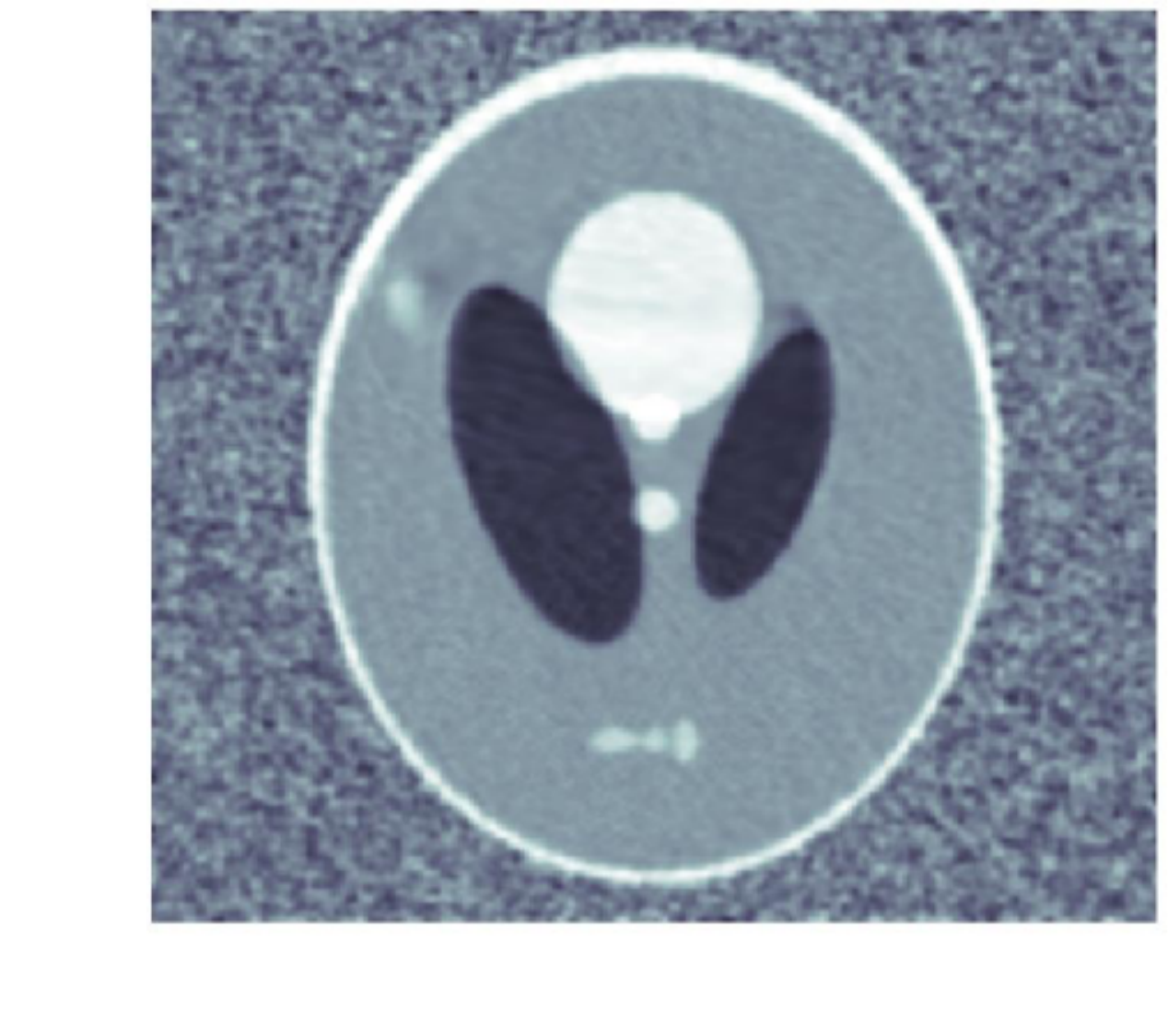}
   \end{minipage}%
   \vskip-0.25\baselineskip
   Photometric trajectory obtained by solving \cref{eq:MetaSpatioTemp} 
   \\[0.75em]
   \begin{minipage}[t]{0.2\textwidth}%
     \centering
     \includegraphics[trim=45 20 5 5, clip, width=\textwidth]{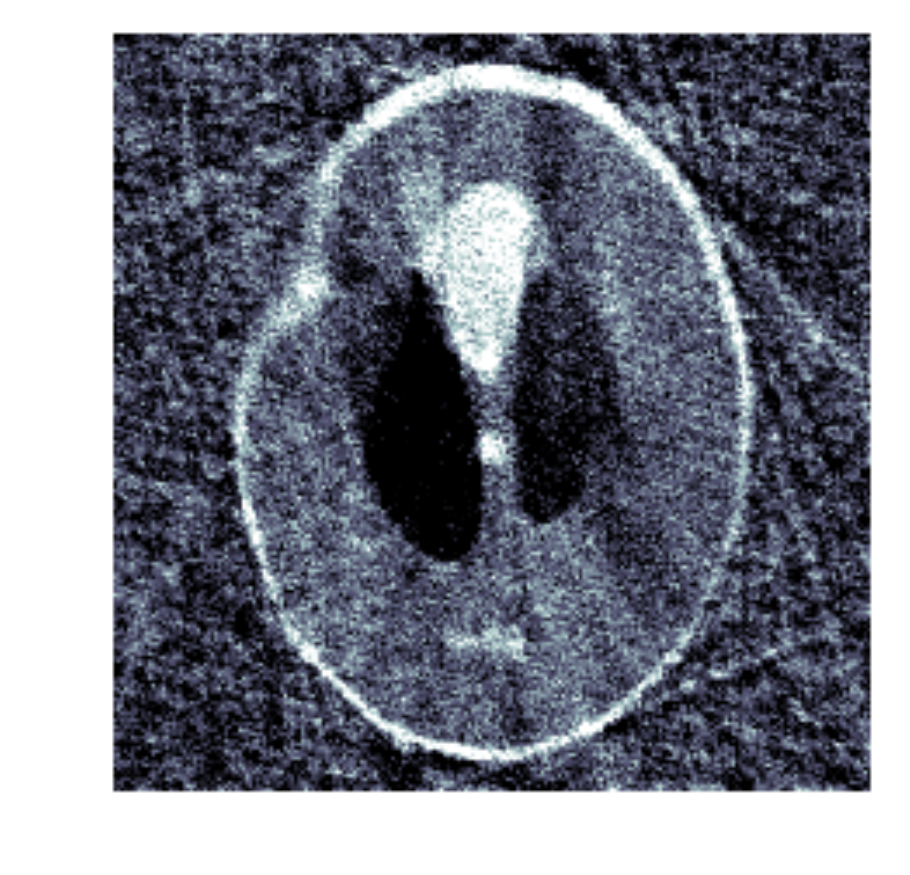}
   \end{minipage}%
   \begin{minipage}[t]{0.2\textwidth}%
     \centering
     \includegraphics[trim=45 20 5 5, clip, width=\textwidth]{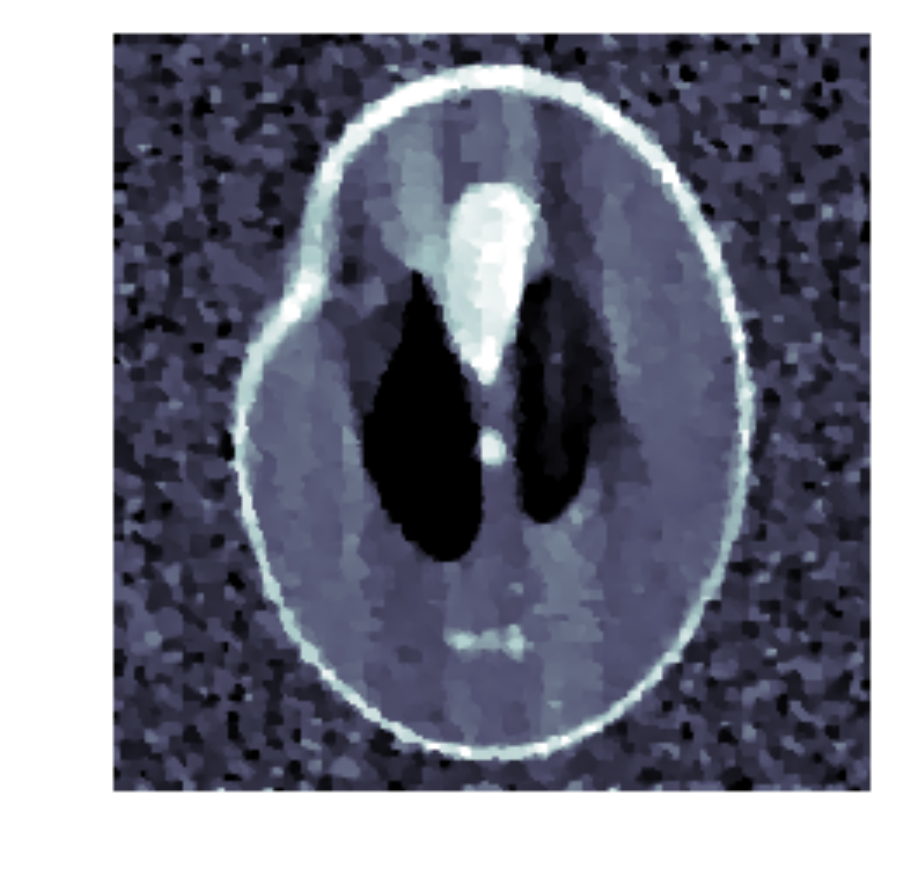}
   \end{minipage}%
   \begin{minipage}[t]{0.23\textwidth}%
     \includegraphics[trim=45 20 5 5, clip, width=\textwidth]{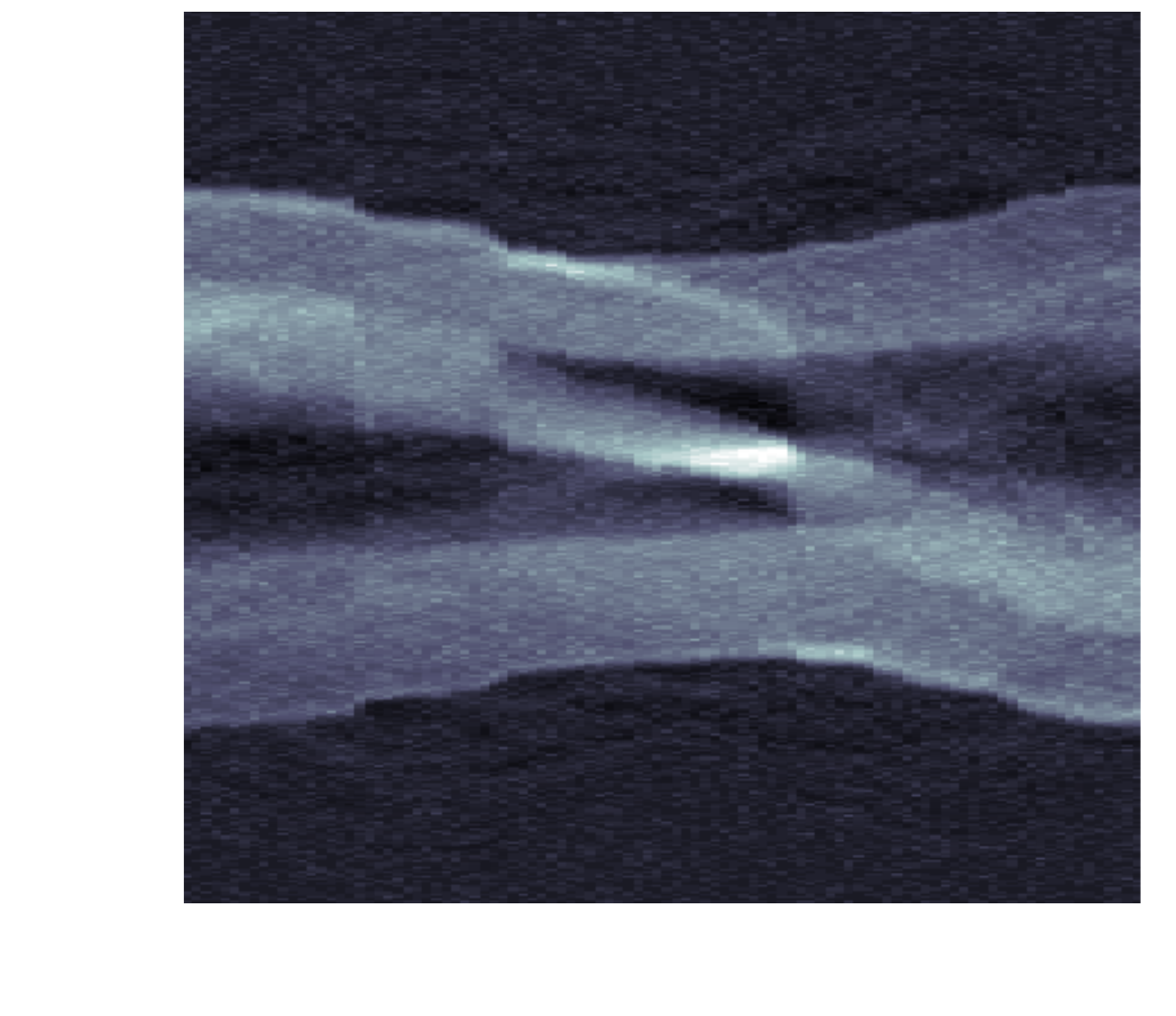}
   \end{minipage}%
   \vskip-0.25\baselineskip
   \Ac{FBP} (left) and \ac{TV} (middle) reconstructions from concatenating the 10 gated data sets (right),
   i.e., sampling the ray transform at 100 angles in $[0, \pi]$.
\caption{Spatiotemporal reconstruction using metamorphosis. 
Top row shows the target image we seek to recover at five (out of 20) selected time points in $[0,1]$.
2nd row shows corresponding gated tomographic data. 
3rd row shows the reconstruction of the target at these time points obtained from \cref{eq:MetaSpatioTemp}.
4:th and 5:th rows show the corresponding shape and photometric trajectories.
Bottom row shows reconstructions assuming a stationary target.}
\label{Fig:TemporalExample}
\end{figure}

\Cref{Fig:TemporalExample} shows results of the above method applied to (gated) 2D tomographic data with a spatiotemporal target image.
We see that \cref{eq:MetaSpatioTemp} can be used for spatiotemporal reconstruction even when (gated) data is highly under sampled and incomplete. 
In particular, one can recover the evolution of the target regarding both shape deformation and photometric changes.
The latter manifests itself in the appearance of the white disc. 

\subsection{Spatiotemporal reconstruction with \ac{LDDMM}}\label{sec:LDDMSpatioTemp}
The aim here is to solve \cref{eq:InvProbDeforTemp} with time continuous data by a variational formulation of the type \cref{eq:SpatioTempDeforTempReg}.
Following \cite{Chen:2019aa}, $\DeforOp_{\deforparam_t} \colon \RecSpace \to \RecSpace$ in \cref{eq:SpatioTempDeforTempReg} (deformation operator) is given by the \ac{LDDMM} framework, so it is parametrised by $\deforparam_t := \vfield(t,\Cdot) \in V$ for some $\vfield \in \VelocitySpace{2}{V}$ as 
\begin{equation}\label{eq:DeforOpLDDMMTemp}
 \DeforOp_{\deforparam_t}(\template) := \diffeoflow{\vfield}{0,t}.\template
 \quad\text{for $\template \in \RecSpace$ and $\diffeoflow{\vfield}{0,t} \in G_V$ as in \cref{eq.FlowDiffeo}.}
\end{equation}
The variant of \cref{eq:SpatioTempDeforTempReg} considered by \cite{Chen:2019aa} is now 
\begin{equation}\label{eq:SpatioTempDeforTempReg2}
  \argmin_{\substack{\template \in \RecSpace \\ t \mapsto \deforparam_t \in \VelocitySpace{2}{V}}}\!\!\!\! 
  \biggl\{ \int_0^T \biggl[ \DataDiscrep\Bigl(\ForwardOp\bigl(t, \DeforOp_{\deforparam_t}(\template) \bigr),\data(t,\Cdot)\Bigr) 
    + \tau \int_0^t \bigl\Vert \deforparam_s \bigr\Vert^2_{V} \der s
    \biggr]\der t
    + \RegFuncSpat_{\gamma}(\template)
  \biggr\}.
\end{equation}
Note that evaluating $\DeforOp_{\deforparam_t}(\template)$ requires solving the \ac{ODE} in \cref{eq.FlowDiffeo}, so \cref{eq:SpatioTempDeforTempReg2} is an \ac{ODE} constrained optimisation problem.

\begin{figure}[htbp]
\centering
\begin{minipage}[t]{\textwidth}%
\centering
   \begin{minipage}[t]{0.2\textwidth}%
     \centering
     \textbf{Gate 1} \\[0.5em]     
     \includegraphics[trim=75 25 60 40, clip, width=\textwidth]{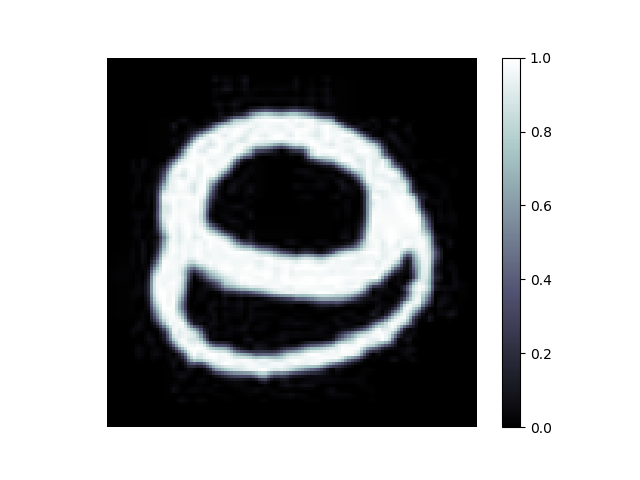}
   \end{minipage}%
   \hfill
   \begin{minipage}[t]{0.2\textwidth}%
     \centering
     \textbf{Gate 2} \\[0.5em]     
    \includegraphics[trim=75 25 60 40, clip, width=\textwidth]{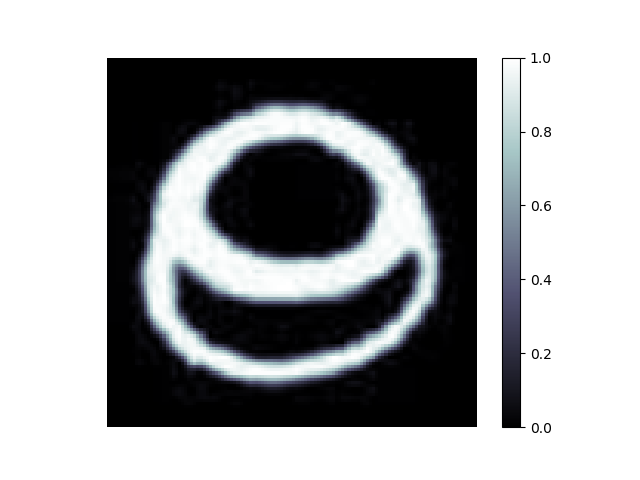}
   \end{minipage}%
   \hfill
   \begin{minipage}[t]{0.2\textwidth}%
     \centering
     \textbf{Gate 3} \\[0.5em]     
     \includegraphics[trim=75 25 60 40, clip, width=\textwidth]{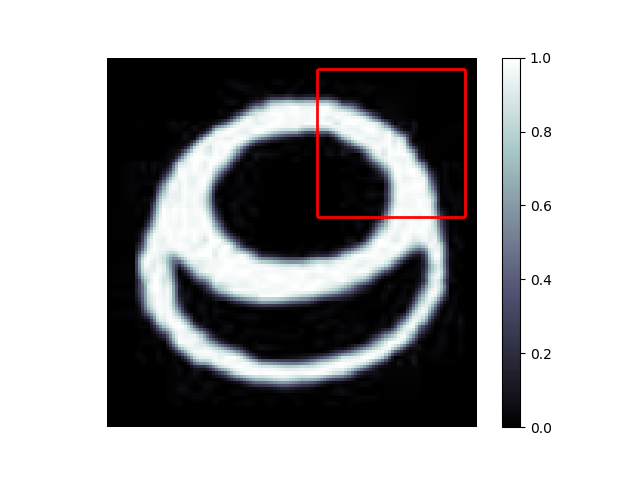}
   \end{minipage}%
      \hfill
   \begin{minipage}[t]{0.2\textwidth}%
     \centering
     \textbf{Gate 4} \\[0.5em]     
     \includegraphics[trim=75 25 60 40, clip, width=\textwidth]{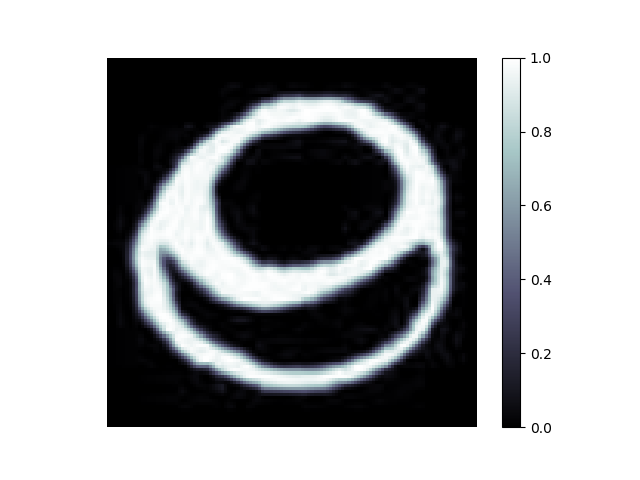}     
   \end{minipage}%
   \\[0.5em]
   Ground truth spatiotemporal image of a heart phantom at four gates.
\end{minipage}%
\\[1em] 
\begin{minipage}[t]{0.2\textwidth}%
     \centering
     \includegraphics[trim=75 25 60 40, clip, width=\textwidth]{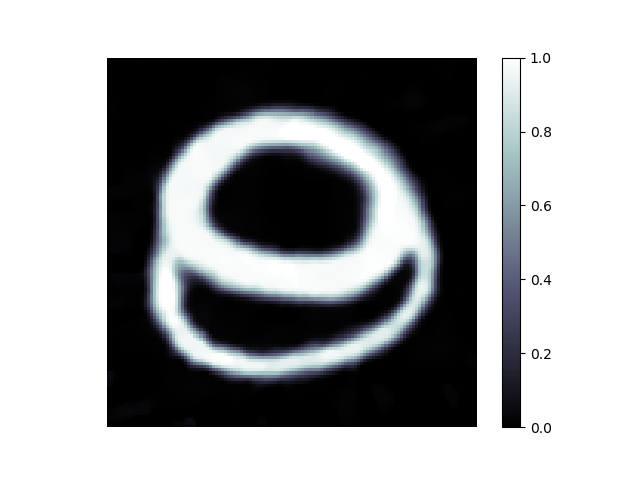}
   \end{minipage}%
   \hfill
   \begin{minipage}[t]{0.2\textwidth}%
     \centering
    \includegraphics[trim=75 25 60 40, clip, width=\textwidth]{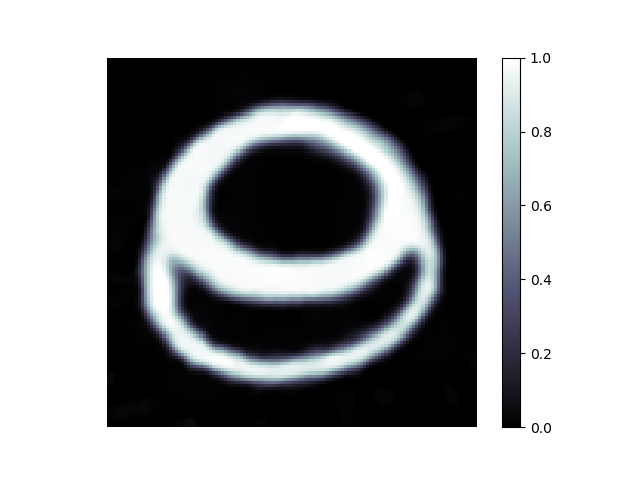}
   \end{minipage}%
   \hfill
   \begin{minipage}[t]{0.2\textwidth}%
     \centering
     \includegraphics[trim=75 25 60 40, clip, width=\textwidth]{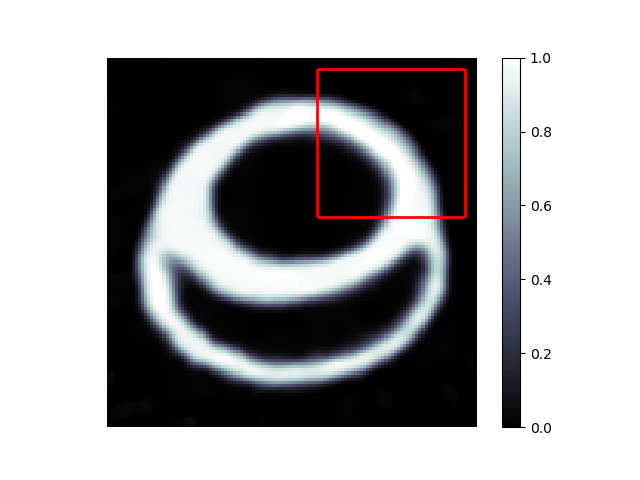}
   \end{minipage}%
   \hfill
   \begin{minipage}[t]{0.2\textwidth}%
     \centering
     \includegraphics[trim=75 25 60 40, clip, width=\textwidth]{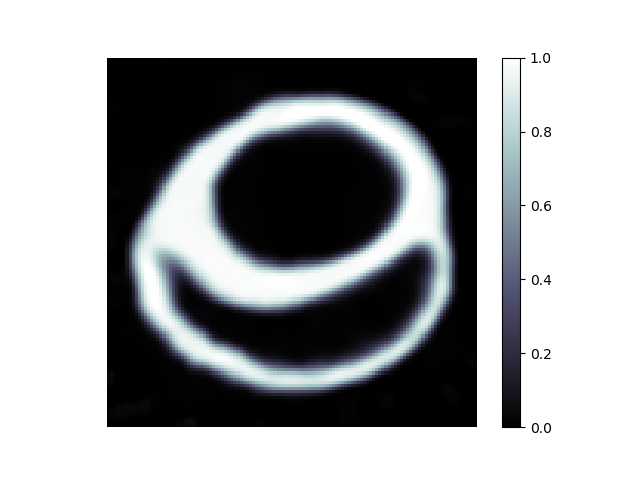}     
   \end{minipage}%
   \\[0.5em]
   \ac{LDDMM} reconstruction of spatiotemporal images from gated tomographic data. 
\\[1em] 
   \begin{minipage}[t]{0.2\textwidth}%
     \centering
     \includegraphics[width=\textwidth]{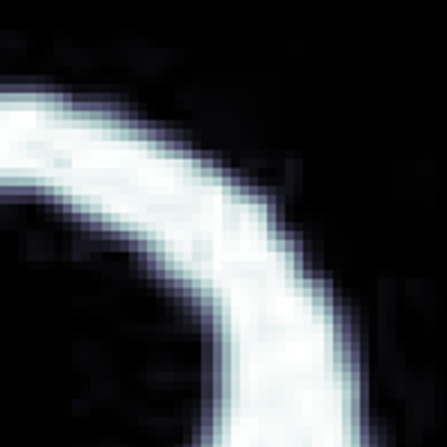}
     \\ ROI in ground truth (gate 3)
   \end{minipage}%
   \hfill
   \begin{minipage}[t]{0.2\textwidth}%
     \centering
     \includegraphics[width=\textwidth]{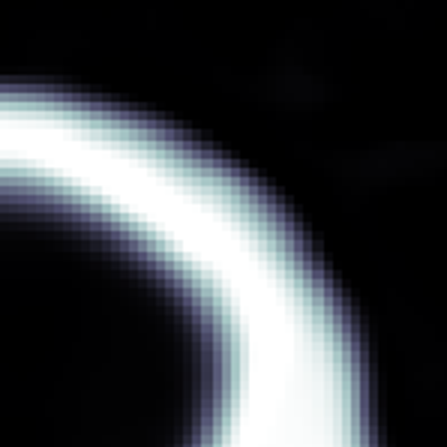}
     \\ ROI in \ac{LDDMM} reco. (gate 3)     
   \end{minipage}%
   \hfill
   \begin{minipage}[t]{0.2\textwidth}%
     \centering
     \includegraphics[width=\textwidth]{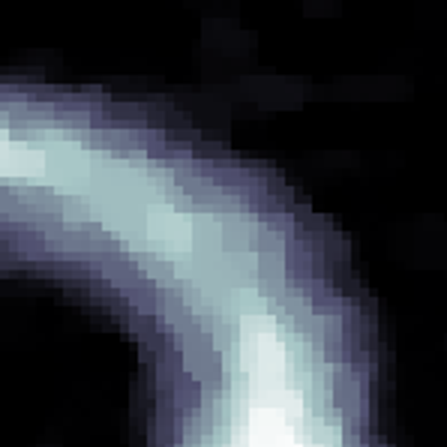}
     \\ ROI in \ac{TV} reconstruction
   \end{minipage}%
   \hfill
   \begin{minipage}[t]{0.2\textwidth}%
     \centering
     \includegraphics[trim=75 25 60 40, clip, width=\textwidth]{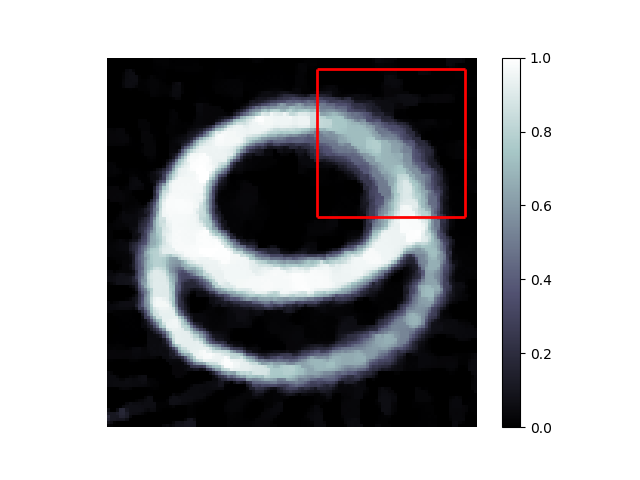}
     \\ \ac{TV} reco. 
   \end{minipage}%
\caption{Spatiotemporal reconstruction using \ac{LDDMM} from gated tomographic data of a heart phantom obtained by solving \cref{eq:SpatioTempDeforTempReg2}. 
The heart phantom is a $120 \times 120$ pixel image with grey-values in $[0, 1]$ that is taken from \cite{GrMi07}.
Data is gated 2D parallel beam tomography where the $i$:th gate has 20 evenly distributed directions in $[(i-1)\pi /5, \pi + (i-1) \pi/5]$. Data (not shown) also has additive Gaussian white noise corresponding to a noise level of about $14.9$dB.
Bottom row compares outcome at an enlarged region of interest (ROI). The ground truth (bottom leftmost image) is compared against \ac{LDDMM} reconstruction (2nd image from left) and \ac{TV} reconstruction (3rd image from left).
The latter is computed assuming a stationary spatiotemporal target and corresponding full image is also shown (bottom rightmost).
It is clear that the cardiac wall is better resolved using a spatiotemporal reconstruction method.
This is essential in \ac{CT} imaging in coronary artery disease.}
\label{Fig:LDDMMSpatioTempExample}
\end{figure}

The temporal regulariser $\RegFuncTemp_{\tau}(t,\Cdot) \colon V \to \Real$ in \cref{eq:SpatioTempDeforTempReg} is given by 
\[ \RegFuncTemp_{\tau}(t,\deforparam) := \tau \int_0^t \bigl\Vert \deforparam_s \bigr\Vert^2_{V} \der s  
   \quad\text{for fixed $\tau>0$,} 
\]
and $\RegFuncSpat_{\gamma} \colon \RecSpace \to \Real$ is the spatial regulariser (typically is of Tikhonov type).
In \cref{Fig:LDDMMSpatioTempExample} we show results from \cite{Chen:2019aa} on using \cref{eq:SpatioTempDeforTempReg2} for spatiotemporal reconstruction in tomography. 

We conclude by pointing out that the model in \cref{eq:SpatioTempDeforTempReg2} can also be stated as \acs{PDE} constrained optimal control problem as shown in \cite[Theorem 3.5]{Chen:2019aa}, see also \cite{Lang:2019aa}.
If $\deforparam_t=\vfield(t,\Cdot) \in V$ for some velocity field $\vfield \in \VelocitySpace{2}{V}$, then \cref{eq:SpatioTempDeforTempReg2} where the deformation operator in \cref{eq:DeforOpLDDMMTemp} is given by the geometric group action in \cref{eq:GeometricGroupAction} is equivalent to
\begin{equation*}
\begin{split}
 &\min_{\substack{\template \in \RecSpace \\ t \mapsto \deforparam_t \in V}} 
 \biggl\{ \int_{0}^{T} \left[\DataDiscrep\Bigl( \ForwardOp_t\bigr(\signal(t,\Cdot)\bigl), \data(t,\Cdot) \Bigr)  
    + \tau \int_{0}^{t} \bigl\Vert \deforparam_s \bigr\Vert^2_{V}\der s
 \right] \der t + \RegFuncSpat_{\gamma}(\template) 
 \biggr\} \\
 & \quad\,\, \text{s.t.  $\partial_t \signal(t, \Cdot) + \bigl\langle \nabla \signal(t,\cdot), \deforparam_t \bigr\rangle_{\Real^n} = 0$.}
\\
 & \quad\,\, \text{\phantom{s.t. }$\signal(0,\Cdot)=\template$.} 
  \end{split}
\end{equation*}  
In a similar manner, if the group action is the mass-preserving as in \cref{eq:MassPreservingGroupAction}, then \cref{eq:SpatioTempDeforTempReg2} becomes 
\begin{equation*}
\begin{split}
 &\min_{\substack{\template \in \RecSpace \\ t \mapsto \deforparam_t \in V}} 
 \biggl\{ \int_{0}^{T} \left[\DataDiscrep\Bigl( \ForwardOp_t\bigl(\signal(t,\Cdot)\bigr), \data(t,\Cdot) \Bigr)  
  +  \tau \int_{0}^{t} \bigl\Vert \deforparam_2 \bigr\Vert^2_{V}\der s \right] \der t + \RegFuncSpat_{\gamma}(\template) 
  \biggr\} \\
 & \quad\,\, \text{s.t. $\partial_t \signal(t, \Cdot) +  \nabla\cdot\bigl(\signal(t, \Cdot)\, \deforparam_t \bigr)= 0$.}
\\
 & \quad\,\, \text{\phantom{s.t. }$\signal(0,\Cdot)=\template$} 
  \end{split}
\end{equation*}  
This establishes the connection between \ac{ODE} based approaches discussed in this section and \ac{PDE} based approaches that are discussed in \cref{sec:PDE}.
As such, it illustrates how one can switch between a reconstruction method based on deformable templates and one based on a motion model (\cref{rem:MotionVsDefor}).

\section{Approaches based on \aclp{PDE}}\label{sec:PDE}
In some applications it is reasonable to assume that the underlying motion is governed by a physical phenomena that can be described by a suitable equation, like a \acs{PDE}. Such an equation can then be used to constrain the motion of the reconstructed target image. 
Focus here is therefore on joint reconstruction and motion estimation as formulated in \cref{eq:SpatioTempConstraintMotModel}.
It has been shown that a joint approach that simultaneously recovers the image sequence and the motion offers a significant advantage over subsequently and separately applying both methods \cite{Burger:2018aa}.

\subsection{Physical motion constraints}
A common model for motion is given by the transport equation
\begin{equation}\label{eqn:convDiffEq}
\begin{cases}
  &\hspace{-1em}\dfrac{\partial \signal}{\partial t}(t,x) + \nabla \cdot \bigl(\vfield(t,x) \signal(t,x)\bigr) =  0, \\[0.75em]
  &\hspace{-1em}\signal(0,x)= \signal_0(x) 
\end{cases}  
\quad\text{for $x \in \signaldomain$ and $t \in [0,T]$.}
\end{equation}
Here, $\signal(t,\Cdot) \colon \signaldomain \to \Real$ is the spatiotemporal image at time $t$ contained in $\RecSpace$ and the velocity field $\vfield(t,x) \colon \signaldomain \to \Real^d$ models the velocity with which points at $x$ move at time $t$. 
The motion model is then given by the underlying equation in \cref{eqn:convDiffEq}, which in turn yields the motion constraint
\begin{equation}\label{eqn:contEq}
  \MotionOp_{\vfield}\bigl( \signal(t,\Cdot) \bigr) := \frac{\partial \signal}{\partial t}(t,\Cdot) + \nabla \cdot \bigl(\vfield(t,\Cdot) \signal(t,\Cdot)\bigr) = 0
  \quad\text{on $\signaldomain \subset \Real^d$.}
\end{equation}
This equation is generally referred to as \emph{continuity equation} and it assumes mass preservation.
Hence, with this model mass can only be continually transformed, no mass can be created, destroyed, or teleported.

A more restrictive model can be directly obtained from \cref{eqn:contEq} under the assumption of incompressible flows, or in our context brightness constancy. 
We give here an alternative derivation, assuming a constant image intensity $\signal(t,x)$ along a trajectory $t \mapsto x(t)$ with velocity $\dot{x}(t)=\vfield(t,x)$, thus we obtain
\begin{equation}\label{eqn:opticalFlowConstraint}
	0 = \frac{\der \signal}{\der t} =  \frac{\partial \signal}{\partial t} + \sum_{i=1}^d \frac{\partial \signal}{\partial x_i}\frac{\der x_i}{\der t} = \partial_t\signal + \nabla \signal\cdot \vfield.
\end{equation}
This equation is also called the \emph{optical flow constraint} and it is a popular approach to model motion between consecutive images \cite{horn1981determining}.
In the following we will base the motion constrained reconstruction as formulated in \cref{eq:SpatioTempConstraintMotModel} on the continuity equation \cref{eqn:contEq}, either assuming mass conservation or the stronger assumption of brightness constancy in form of the optical flow model. 
For both models, the time dependent parametrisation of the motion model is by velocity fields, i.e., the motion model is given as $\Psi_{\motionparam_t}\bigl( \signal(t,\Cdot) \bigr)$ where $\motionparam_t := \vfield(t,\Cdot)$ for some sufficiently regular velocity field $\vfield \colon [0,T] \times \signaldomain \to \Real^d$ (motion field). 
Henceforth, we use the notation $\MotionOp_{\vfield} := \Psi_{\motionparam_t}$.

\subsubsection{Joint motion estimation and reconstruction}
A joint model for motion estimation and tomographic reconstruction can, based on the motion constrained model in \cref{eq:SpatioTempConstraintMotModel}, be formulated for $p\in\{1,2\}$ and $q,r > 1$ as 
\begin{equation}\label{eqn:JointModel}
\begin{split}
  \argmin_{\substack{t \mapsto \signal(t,\Cdot) \in \RecSpace \\ t \mapsto \vfield(t,\Cdot)\in V}}
    & \int_0^T \left[ \frac{1}{p}\Bigl\| \ForwardOp\bigl(t, \signal(t,\Cdot) \bigr) - \data(t,\Cdot) \Bigr\|_p^p 
         + \alpha \bigl| \signal(t,\Cdot) \bigr|_{\text{BV}}^q + \beta \bigl| \vfield(t,\Cdot) \bigr|_{\text{BV}}^r \right] \der t,\\
  \text{s.t. } 
    & \MotionOp_{\vfield}\bigl( \signal(t,\Cdot) \bigr) = 0 \text{ on $\signaldomain \subset \Real^d$.}
\end{split}
\end{equation} 
Here we use for both image sequence and vector field the respective total variation as a regulariser, given by the semi-norm in the space of bounded variation. Consequently, given fixed domain $\signaldomain \subset \Real^d$, the spaces under consideration here are $\RecSpace=\BVSpace(\signaldomain,\Real)$ for the reconstructions and $V=\BVSpace(\signaldomain,\Real^d)$ for the corresponding vector field. Other models can be considered such as $L^2$-regulariser for the mass conservation or other convex regulariser, see \cite{Burger:2018aa,dirks2015variational} for details. We furthermore assume the forward operator $\ForwardOp(t,\Cdot) \colon \RecSpace \to \DataSpace$ to be a bounded linear operator to some Hilbert space $\DataSpace$. In particular, it can be time dependent \cite{Burger:2017aa,frerking2016variational}.

The motion constraint in \cref{eqn:contEq} is used to describe how image sequence and vector fields are connected. From the perspective of tomographic reconstructions, the motion constraint acts as an additional temporal regulariser along the motion field $\vfield$. Instead of imposing the motion constraint exactly as in \cref{eqn:JointModel} we can also relax it and add as a least-squares term to the functional itself, cf. \cite{Burger:2018aa}.

In order to establish existence of minimisers of \cref{eqn:JointModel}, we need ensure appropriate weak-star compactness of sublevel sets and lower semicontinuity. We will restrict the following results here now to dimension $d=2$. 
For the minimisation we consider the space
\begin{multline}\label{eq.Dset}
  D := \Bigl\{ (\signal,\vfield) \in L^{\min\{p,q\}}\bigl([0,T];\RecSpace \bigr) \times L^r\left([0,T];V\right)
    \mid
    \\
    \| \vfield \|_{\infty} \le c_v < \infty \text{ and } \| \nabla \cdot \vfield  \|_{E} \leq c_d 
    \Bigr\},
\end{multline}
where $E$ is a Banach space continuously embedded into $L^m([0,T];L^k(\signaldomain,\Real^d))$,
$k >p$ and $m > q^*$ with $q^*$ being the H\"older conjugate of $p$. 
We can now state an existence result for the joint model \cref{eqn:JointModel} that is proved in \cite{Burger:2018aa}.
\begin{theorem}{(Existence of minimisers to \cref{eqn:JointModel})}
Given a linear forward operator $\ForwardOp(t,\Cdot):\RecSpace\to\DataSpace$, $p\in\{1,2\}$ and dimension $d=2$, let $1<q,r$ and
\[
\Op{J}(\signal,\vfield) :=  \int_0^T \Bigl[ \frac{1}{p}\left\| \ForwardOp\bigl(t, \signal(t,\Cdot) \bigr) - \data(t,\Cdot) \right\|_p^p + \alpha |\signal(t,\Cdot)|_{\text{BV}}^q + \beta | \vfield(t,\Cdot)|_{\text{BV}}^r \Bigr]\der t.
\]
Furthermore, let $\ForwardOp$ be such that it does not eliminate constants, i.e. $\ForwardOp(t,\vec{1}) \neq 0$ for all $t \in [0,1]$.
Then, there exists a minimiser of $\Op{J}(\signal,\vfield)$ in the constraint set
\[
S := \bigl\{ (\signal,\vfield) \in D \mid  \MotionOp_{\vfield}(\signal) = 0 \bigr\}
\quad\text{where $D$ is given as in \cref{eq.Dset}.}
\]
\end{theorem}
The proof for $p=2$ follows from \cite{dirks2015variational,Burger:2018aa} and the case for $p=1$ follows similar arguments as outlined in \cite{frerking2016variational}. Existence for the unconstrained case is proved by incorporating the constraint as a penalty term in the functional $\Op{J}$ as shown in \cite{Burger:2018aa}. We note here that the choice $q,r > 1$ has to be made in the analysis in order to avoid dealing with measures in time. 
In the computational use cases considered below, it is however reasonable to set $q=r=1$.

\subsubsection{Implementation and reconstruction}
For computational reasons, as well as to allow slight deviations from the motion model, it is advantageous to consider a penalised version instead of the constrained formulation \cref{eqn:JointModel}. Then the joint minimisation problem for spatiotemporal reconstructions can be written as
\begin{multline}\label{eqn:penalJointModel}
   \argmin_{\substack{t \mapsto \signal(t,\Cdot) \in \RecSpace \\ t \mapsto \vfield(t,\Cdot)\in V}} \int_0^T \Bigl[ \frac{1}{p}\left\|\ForwardOp\bigl(t, \signal(t,\Cdot)\bigr) - \data(t,\Cdot) \right\|_p^p 
  \\
    + \alpha | \signal(t,\Cdot) |_{\text{BV}} 
    + \gamma \left\| \MotionOp_{\vfield}\bigl(\signal(t,\Cdot)\bigr) \right\|_1 + \beta | \vfield(t,\Cdot) |_{\text{BV}} \Bigr] \der t,
\end{multline}
where convergence to the constrained model is given for $\gamma\to\infty$. In practice, the BV-semi-norm is replaced by the discrete isotropic total variation. 

\begin{figure}[htbp]
\centering
\begin{minipage}[t]{\textwidth}%
\centering
   \begin{minipage}[t]{0.33\linewidth}%
     \centering
     \includegraphics[width=\linewidth]{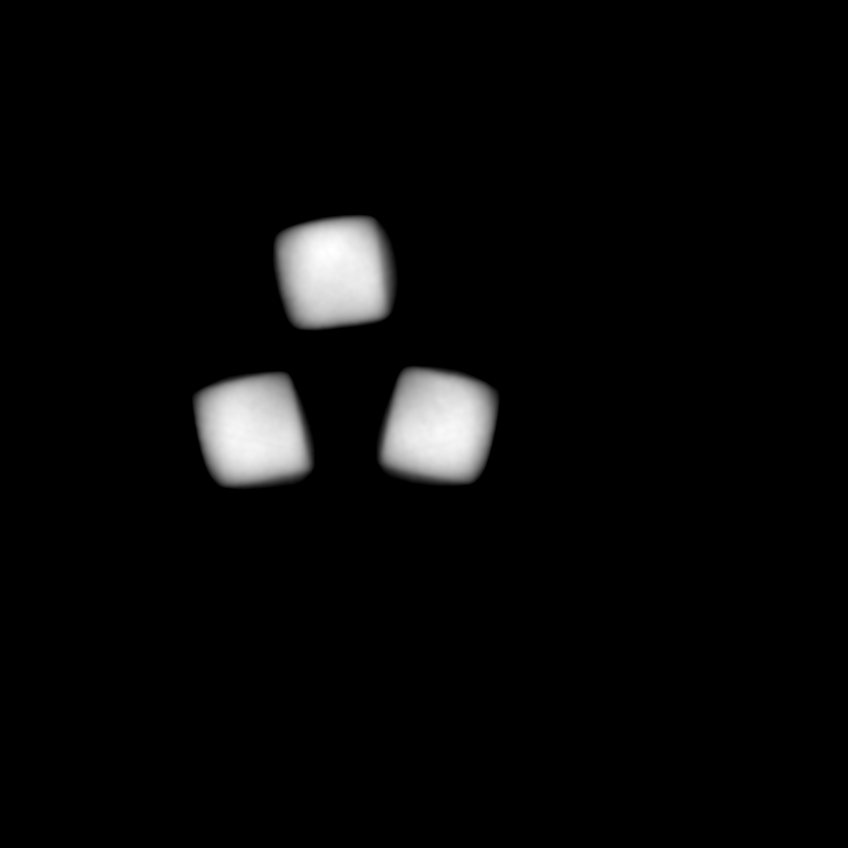}
   \end{minipage}%
   \hfill
   \begin{minipage}[t]{0.33\linewidth}%
     \centering
    \includegraphics[width=\linewidth]{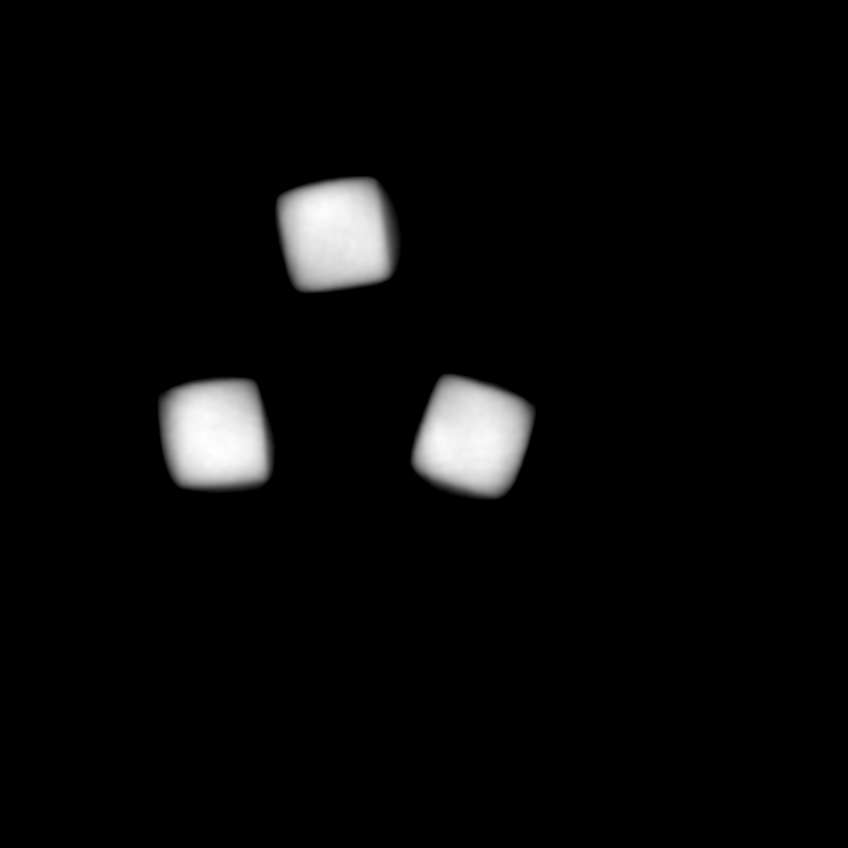}
   \end{minipage}%
   \hfill
   \begin{minipage}[t]{0.33\linewidth}%
     \centering
     \includegraphics[width=\linewidth]{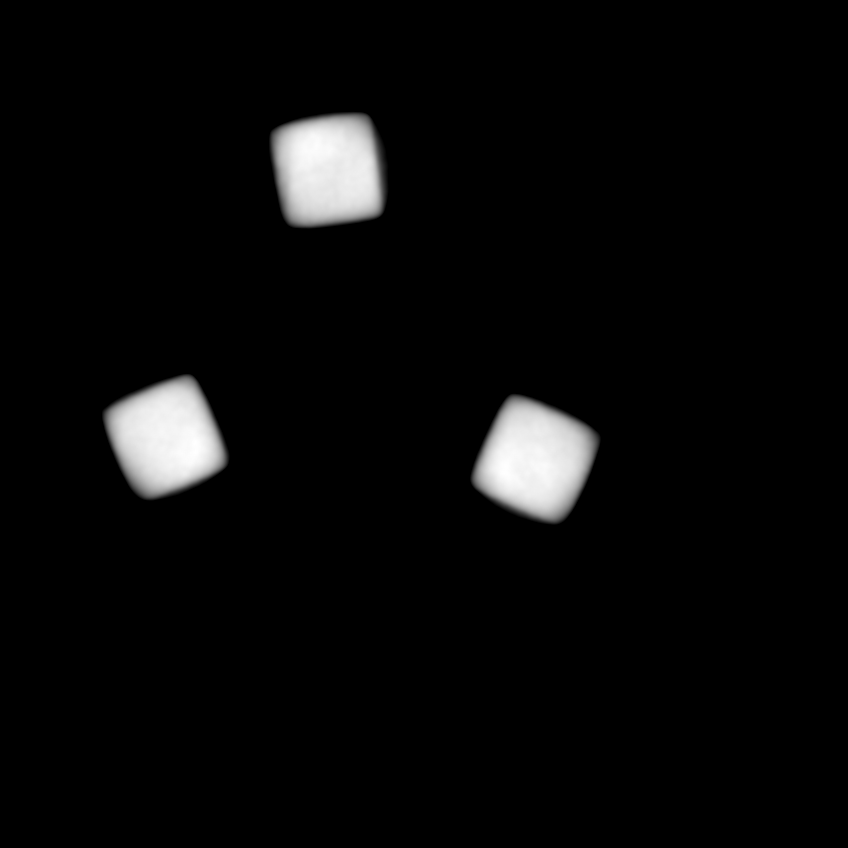}
   \end{minipage}%
   \\[0.5em]
   Ground truth spatiotemporal image at three time steps 7, 18, 25 out of 30.
\end{minipage}%
\\[1em] 
\begin{minipage}[t]{\textwidth}%
   \begin{minipage}[t]{0.33\linewidth}%
     \centering
     \textbf{Single angular sampling} \\[0.5em]          
     \includegraphics[height=54pt]{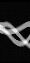}
     \includegraphics[height=54pt]{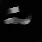}     
   \end{minipage}%
   \hfill
   \begin{minipage}[t]{0.33\linewidth}%
     \centering
     \textbf{Double angular sampling} \\[0.5em]          
     \includegraphics[height=54pt]{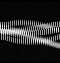}
     \includegraphics[height=54pt]{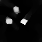}     
   \end{minipage}%
   \hfill
   \begin{minipage}[t]{0.33\linewidth}%
     \centering
     \textbf{Random sampling} \\[0.5em]               
     \includegraphics[height=54pt]{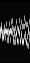}
     \includegraphics[height=54pt]{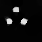}     
   \end{minipage}%
   \\[0.5em]
   Reconstructions and data from two consecutive angular sampling schemes with one and two source-detector pairs (left and middle) and a sampling scheme with only one measurement at each time instance from a randomly (uniformly) chosen direction (right). The data over time is shown to the left and reconstructions for time point 18 are shown to the right.
\end{minipage}%
\caption{Reconstructions from \cite{Burger:2017aa} of experimental X-ray data using the approach in \cref{eqn:penalJointModel} with an optical flow constraint.
Top row shows the ground truth spatiotemporal image and bottom row shows data and reconstruction for three sampling schemes.}\label{fig:stonesGT}
\end{figure}

\begin{figure}[htbp]
\centering
\begin{minipage}[t]{0.49\textwidth}%
\centering
\textbf{$L^1$ fidelity term} \\[0.5em]          
   \begin{minipage}[t]{0.49\linewidth}%
     \centering
     \includegraphics[width=\linewidth]{figures/data_stones/XS_T30_L1_1RandMeasAng/result_data_30Time_1RandEach_18_rec.png}
     \\
     \includegraphics[width=\linewidth]{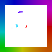}     
   \end{minipage}%
   \hfill
   \begin{minipage}[t]{0.49\linewidth}%
     \centering
    \includegraphics[width=\linewidth]{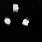}
     \\
     \includegraphics[width=\linewidth]{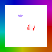}     
   \end{minipage}%
\end{minipage}%
\hfill
\begin{minipage}[t]{0.49\textwidth}%
\centering
\textbf{$L^2$ fidelity term} \\[0.5em]          
   \begin{minipage}[t]{0.49\linewidth}%
     \centering
     \includegraphics[width=\linewidth]{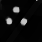}
     \\
     \includegraphics[width=\linewidth]{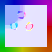}     
   \end{minipage}%
   \hfill
   \begin{minipage}[t]{0.49\linewidth}%
     \centering
    \includegraphics[width=\linewidth]{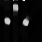}
     \\
     \includegraphics[width=\linewidth]{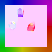}     
   \end{minipage}%
\end{minipage}%
\caption{Reconstruction results for the random sampling with both $p=1,2$ for the fidelity term in \cref{eqn:penalJointModel} for time points 17 and 25. The left images show that $L^1$-norm clearly favours sparse reconstructions with a resulting sparse motion field. In contrast, the $L^2$-norm shown in the right favours smoother reconstructions and motion fields.}
\label{fig:stonesL1vsL2}
\end{figure}

As the penalised formulation depends on the motion model $\MotionOp_{\vfield}(\signal)$, the energy to be minimised is nonlinear and therefore non-convex. Additionally, it is non-differentiable due to the involved $L^1$-norms and hence the computation of a solution to \cref{eqn:penalJointModel} is numerically challenging. Thus, in practice it is advised to compute solutions using an intertwined scheme, which means that we split the joint model into two alternating optimisation problems, one for $\signal$ and the other for $\vfield$:
\begin{align}
\label{eqn:jointSpatial}
\signal^{k+1} = \argmin_{t \mapsto \signal(t,\Cdot) \in \RecSpace} 
  &\int_0^T \Bigl[ \frac{1}{p}\left\|\ForwardOp(t,\signal)-\data \right\|_p^p  + \alpha | \signal |_{\text{BV}} 
    + \gamma \left\| \Psi_{\vfield^k}(\signal)\right\|_1 \Bigr] \der t \\
\label{eqn:jointMotion}
\vfield^{k+1} = \argmin_{t \mapsto \vfield(t,\Cdot)\in V} 
  &\int_0^T  \Bigl[ \left\| \MotionOp_{\vfield}(\signal^{k+1})\right\|_1 
    + \frac{\beta}{\gamma} | \vfield |_{\text{BV}} \Bigr] \der t.
\end{align}
Most importantly, both subproblems are now linear and convex, but we note that the solution of the alternating scheme might correspond to a local minima of the joint model. In practice, one would initialise $\signal^0=0$ and $ \vfield = \vec{0}$, then the first minimisation problem for $\signal^1$ corresponds to a classic total variation regularised solution for each image time instance separately followed by a motion estimation. Reconstructions from \cite{Burger:2017aa} using this alternating scheme for experimental $\mu$\ac{CT} data are shown in Figure \ref{fig:stonesGT} and an illustration of the influence of $L^p$-norms in the data fidelity in Figure \ref{fig:stonesL1vsL2}.

One can use any optimisation algorithm that supports non-differentiable terms for computing solutions to each of the subproblems \cref{eqn:jointSpatial,eqn:jointMotion}.  In dimension $d=2$ one could simply use a primal-dual hybrid gradient scheme \cite{chambolle2011first} as outlined in \cite{Burger:2017aa}, see also \cite{aviles2018compressed}, here both applications use the optical flow constraint \cref{eqn:opticalFlowConstraint}. 
In higher dimensions where the computational burden of the forward operator becomes more prevalent, it is advised to consider other schemes with fewer operator evaluations, we refer to \cite{lucka2018enhancing} for an application to dynamic 3D photoacoustic tomography as well as \cite{djurabekova2019application} for dynamic 3D computed tomography. 

To conclude this section, we mention that in other applications it might be more suitable to  require mass conversation using the continuity equation instead, see for instance \cite{lang2019joint}.

\section{Data driven approaches}\label{sec:ML}
The variational approaches outlined in \cref{sec:tempindependentrecon,sec:MotionRec,sec:DeforTemp} come with \emph{two serious drawbacks} that limit their applicability. 
First, they typically result in complex non-convex optimisation problems that are difficult to solve reasonably fast in time-critical applications.
Second, they rely on a handcrafted family of parametrised temporal models that need to be computationally feasible, yet expressive enough to represent relevant temporal evolution. 

Data driven models, and especially those based on deep learning, offer means to address these drawbacks.
Once trained, a deep learning model is typically very fast to apply. 
Next, its large model capacity also allows for capturing complicated temporal evolution that is otherwise difficult to account for in handcrafted models.
Embedding a deep learning model into a spatiotemporal reconstruction method is however far from straightforward. 

\Cref{sec:DataDrivenDirect} outlines how to do this in the context of the reconstruction method in \cref{sec:tempindependentrecon}.  
The situation is more complicated for reconstruction methods that use explicit temporal models.
These methods rely on joint optimisation of the image and the temporal model, so the latter needs to be parametrised.
Embedding a deep learning based temporal model is therefore only feasible if said parametrisation is preserved and most existing deep learning approaches for temporal modelling of images do not fulfil this requirement.
\Cref{sec:LearnedDeforOp} surveys selected deep learning models for deformations that can be embedded into reconstruction methods that use a deformable template (\cref{sec:DeforTemp}).
Finally, \cref{sec:LearnedMotionModel} considers embedding deep learning based models into reconstruction methods that use motion models (\cref{sec:MotionRec}).

\subsection{Data driven reconstruction without temporal modelling}\label{sec:DataDrivenDirect}
A data driven approach for solving \cref{eq:InvProb} starts by considering a family $\{ \RecOp_{\MLparam} \}_{\MLparam \in \MLparamSet}$ of reconstruction operators $\RecOp_{\MLparam}(t,\cdot) \colon \DataSpace \to \RecSpace$.
In deep learning, $\RecOp_{\MLparam}$ is represented by a deep neural network with network parameters $\MLparam$.
The learning amounts to finding the reconstruction operator $\RecOp_{\MLparamOpt}(t,\cdot) \colon \DataSpace \to \RecSpace$ where $\MLparamOpt\in \MLparamSet$ is learned from (supervised) training data as 
\begin{equation}\label{eqn:learningTask}
  \MLparamOpt \in \argmin_{\MLparam \in \MLparamSet} L(\MLparam)
  \quad\text{where}\quad
  L(\MLparam) := \sum_{i=1}^N \int_0^T \ell_{\RecSpace}\Bigl( \RecOp_{\MLparam}\bigl(t,\data_i(t,\cdot)\bigr), \signal_i(t,\cdot) \Bigr) \der t.
\end{equation}
Here, $\ell_{\RecSpace} \colon \RecSpace \times \RecSpace \to \Real$ quantifies goodness-of-fit of images and $t \mapsto \data_i(t,\cdot) \in \DataSpace$ and $t \mapsto \signal_i(t,\cdot) \in \RecSpace$ for $i=1,\ldots,N$ represent ground truth spatiotemporal image and corresponding noisy data, i.e.,
\begin{equation}\label{eq:TrainingData}
 t \mapsto (\signal_i(t,\Cdot),\data_i(t,\Cdot)\bigr) \in \RecSpace \times \DataSpace 
   \quad\text{satisfy \cref{eq:InvProb} for $i=1,\ldots,N$.}
\end{equation}

A key component is to specify the appropriate (deep) neural network architecture for $\RecOp_{\MLparam}(t,\cdot) \colon \DataSpace \to \RecSpace$.
One option is to set $\RecOp_{\MLparam} := \Op{P}_{\MLparam} \circ \Op{A}^{\dagger}$ where $\Op{A}^{\dagger}(t,\cdot) \colon \DataSpace \to \RecSpace$ is a (non-learned) reconstruction operator for solving \cref{eq:InvProb} and $\Op{P}_{\MLparam}(t,\cdot) \colon \RecSpace \to \RecSpace$ is a data driven post-processing operator \cite{hauptmann2019real,kofler2019spatio}.
Hence, the input to the data driven part is a spatiotemporal image and the output is an `improved' spatiotemporal image.
Such a model is trained against supervised data consisting of pairs of spatiotemporal images, one representing ground truth and the other the output from said reconstruction method. 
Alternatively, one can learn an updates in an unrolled iterative scheme that is derived from some fixed point-scheme for solving \cref{eq:DirectSpatioTempReg} as in \cite{schlemper2017deep}.
This includes a handcrafted forward operator, which in \cite{schlemper2017deep} is time independent (Fourier transform) but its sampling in $\datadomain$ depends on time. 
Such an approach needs supervised training data of the form \cref{eq:TrainingData} for its training. 

Common for both approaches is that the neural network architecture does not make use of any explicit deformation/motion model.
As such, they represent data driven variants of methods outlined in \cref{sec:tempindependentrecon}.

\subsection{Learning deformation operators}\label{sec:LearnedDeforOp}
Focus here is on using a deep learning model in a reconstruction method that uses a deformable template (\cref{sec:DeforTemp}).
One possibility is to use deep learning to model the time evolution $t \mapsto \deforparam_t$ of the deformation parameter, which is the approach (deep diffeomorphic normalising flow) taken in  \cite{Salman:2018aa}. 
Another option is to use possibility in defining the parametrised deformation operator $\DeforOp_{\deforparam_t} \colon \RecSpace \to \RecSpace$ in \cref{eq:DeforTemp}.
Our emphasis is on the latter, which essentially amounts to considering deep learning approaches for image registration.

There is a rich theory of variational approaches to image registration, see the books \cite{GrMi07,Younes:2019aa} and surveys in \cite{Pennec:2020aa,Kushnarev:2020aa}.
The common trait with these approaches is that deformation models are parametrised. 
A variational problem is then formulated to select the `best' deformation by regularising the deformation itself to avoid overfitting while ensuring adequate match between the template and target images.
Recently, there are also many publications that consider deep learning for image registration, see \cite{Shen:2017aa,Litjens:2017aa,Fu:2019aa,Haskins:2020aa} for surveys. 
Most of these learn a deformation operator directly from pairs of template and target images without accounting for any specific parametrisation, i.e., the learned deformation operator is not parametrised by a deformation parameter\footnote{The temporal model is defined by considering a time dependent deformation parameter. The deep neural network representing the deformation operator also has parameters, but these are not the same as the deformation parameter. In particular, the network parameters are set during training. In contrast, the deformation parameter varies with time.}.

A key aspect is that the trained deep neural network is parametrised explicitly with a (deformation) parameter and it does not require re-training when the (deformation) parameter changes.
Such a data driven model can be used in reconstruction with deformable templates as shown in \cite{Liu:2019aa,Pouchol:2019aa} for the case when data is time discretised.
Both these approaches start out by stating a variational model of the type \cref{eq:SpatioTempDeforTempReg}, which is then solved using an intertwined approach of the type \cref{eq:SpatioTempRecMatchedTrajectory}. 
Here one considers diffeomorphic deformations as defined by the \ac{LDDMM} framework, i.e., deformation operators are parametrised as in  \cref{eq:DeforOpLDDMM}.
A key part is the usage of deep learning based deformation operators that are of the same form, i.e., the trained deep neural network retains the parametrisation in \cref{eq:DeforOpLDDMMTemp}. 
\emph{In the following, our emphasis is on deep learning models for registration that adhere to a specific pre-defined parametrisation.}
Stated more precisely, one seeks to use a data driven model for these  deformation operator that belongs to a pre-defined parametrised family $\{ \DeforOp_{\deforparam} \}_{\deforparam \in \DeforParamSet}$.

One way to achieve the above is by learning a mapping $\Lambda_{\MLparam} \colon \RecSpace \times \RecSpace \to \DeforParamSet$ that predicts the deformation parameter necessary for deforming a template to a target as  
\[ 
   \deforparam := \Lambda_{\MLparam}(\template,\target)
   \implies 
   \DeforOp_{\deforparam}(\template) \approx \target 
   \quad\text{for $\template,\target \in \RecSpace$.}
\]
Note here that $\MLparam \in \MLparamSet$ is the deep neural network parameter that is set during training. It is \emph{not} the same as the deformation parameter $\deforparam \in \DeforParamSet$, which parametrises the deformation operator $\DeforOp_{\deforparam} \colon \RecSpace \to \RecSpace$ and which is a control variable in the variational approaches for reconstruction. 
In some sense, $\Lambda_{\MLparam}$ can be seen as a generative model for the deformation parameter.

The mapping $\Lambda_{\MLparam} \colon \RecSpace \times \RecSpace \to \DeforParamSet$ can be trained in an unsupervised setting given access to sufficient amount of training data of the form  
\begin{equation}\label{eq:UnSupervisedDataImageReg}
 (\target^i,\template^i) \in \RecSpace \times \RecSpace 
 \quad\text{for $i=1,\ldots, N$}
\end{equation}
by computing $\MLparamOpt \in \MLparamSet$ as
\begin{equation}\label{eq:LossRegUnSupervised}
   \MLparamOpt \in \argmin_{\MLparam \in \MLparamSet} L(\MLparam)
   \quad\text{where}\quad
   L(\MLparam) := \sum_{i=1}^N \ell_{\RecSpace}\Bigl( \DeforOp_{\Lambda_{\MLparam}(\template^i, \target^i)}(\template^i),\target^i\Bigr).
\end{equation}  
Here, $\ell_{\RecSpace} \colon \RecSpace \times \RecSpace \to \Real$ is a distance notion between images, e.g., the squared $L^2$-norm if $\RecSpace = L^2(\signaldomain)$. 
One can also add an additional regularisation term to \cref{eq:LossRegUnSupervised} that measures registration accuracy in the image space $\RecSpace$.
\begin{remark}
One can also train $\Lambda_{\MLparam} \colon \RecSpace \times \RecSpace \to \DeforParamSet$ in an supervised setting assuming access to training data of the form
\begin{equation}\label{eq:SupervisedDataImageReg}
 (\target^i,\template^i,\deforparam^i) \in \RecSpace \times \RecSpace \times  \DeforParamSet
 \quad\text{where $\target^i \approx \DeforOp_{\deforparam^i}(\template^i)$ for $i=1,\ldots, N$.}
\end{equation}
The network parameter $\MLparam \in \MLparamSet$ is trained against the supervised data in \cref{eq:SupervisedDataImageReg} by computing $\MLparamOpt \in \MLparamSet$ as 
\begin{equation}\label{eq:LossRegSupervised}
 \MLparamOpt \in \argmin_{\MLparam \in \MLparamSet} L(\MLparam)
   \quad\text{where}\quad
   L(\MLparam) := \sum_{i=1}^N \ell_{\DeforParamSet}\bigl( \Lambda_{\MLparam}(\template^i, \target^i), \deforparam^i \bigr)
\end{equation}  
Here, $\ell_{\DeforParamSet} \colon \DeforParamSet \times \DeforParamSet \to \Real$ is a distance notion between deformation parameters, so $\DeforParamSet$ must have a metric space structure.
Hence, the registration accuracy is measured in the deformation parameter set $\DeforParamSet$.
\end{remark}

An example of this approach is Quicksilver \cite{yang2017quicksilver}, which considers deformation operators $\{ \DeforOp_{\deforparam} \}_{\deforparam}$ given by the \ac{LDDMM} framework.
Then, $\deforparam := \vfield(1,\Cdot)$ for some velocity field $\vfield \colon [0,1] \times \signaldomain \to \Real^d$ and 
\begin{equation}\label{eq:DeforOpLDDMM}
 \DeforOp_{\deforparam}(\template) := \diffeoflow{\vfield}{0,1}.\template
 \quad\text{with $\diffeoflow{\vfield}{0,1} \in G_V$ as in \cref{eq.FlowDiffeo},}
\end{equation}
and the group action is typically geometric \cref{eq:GeometricGroupAction} or mass preserving \cref{eq:MassPreservingGroupAction}.
It is known that the vector field $\deforparam \in \DeforParamSet$ that registers a template to a target can be computed by geodesic shooting, see \cite{Miller:2006aa} and \cite[Section~10.6.4]{Younes:2019aa}.
The registration problem, which is to find $\deforparam$, thus reduces to finding the initial momenta.
Quicksilver \cite{yang2017quicksilver} trains a deep neural network in the unsupervised setting (as in \cref{eq:LossRegUnSupervised}) to learn these initial momenta. 
The network architecture for $\Lambda_{\MLparam} \colon \RecSpace \times \RecSpace \to \DeforParamSet$ is of \ac{CNN} type with an encoder and a decoder. 
The encoder acts as a feature extraction for both template and target images. 
The extracted features are then concatenated and fed into the decoder, which consists of three independent convolutional networks that predict the momenta for the three dimensions. 
To recover from prediction errors, a correction networks with the same architecture is used for predicting the prediction error. 
Training such a deep neural network model with entire images is challenging, so Quicksilver only uses patches of images as input. 
In this way, relatively few images and ground truth momenta result in a large amount of training data. 
A drawback is that the patches are extracted from the target, template and deformation are on the same spatial grid locations, so the deformed patch in the target is assumed to lie (predominantly) in the same location as the one in the template image.
This assumes the deformation is relatively small.

Another similar approach is VoxelMorph \cite{Balakrishnan:2019aa} where training is performed in an unsupervised manner (as in \cref{eq:LossRegUnSupervised}) with only pairs of template and morphed image. The output is the displacement field $\deforparam \in \DeforParamSet$ necessary to register a template against a target, e.g., using an \ac{LDDMM} based deformation operator.
VoxelMorph uses \ac{CNN} architecture similar to U-net for $\Lambda_{\MLparam} \colon \RecSpace \times \RecSpace \to \DeforParamSet$ that consists of encoder and decoder sections with skip connections.
The unsupervised loss \cref{eq:LossRegUnSupervised} can be complemented by an auxiliary loss that leverages anatomical segmentations at training time.
The trained network can also provide the registered image, i.e., it offers a deep learning based registration operator.
A further development of VoxelMorph is FAIM \cite{Kuang:2018aa} that has fewer trainable parameters (i.e., dimension of $\MLparam$ in FIAM is smaller than the one in VoxelMorph). 
Authors also claim that FAIM achieves higher registration accuracy than VoxelMorph, e.g., it produces deformations with many fewer `foldings', i.e. regions of non-invertibility where the surface folds over itself.

One may also learn the spatially-adaptive regulariser that is used for defining the deformation operator \cite{Niethammer:2019aa}. See also \cite{Mussabayeva:2019aa} for a closely related approach where one learns the regulariser in the \ac{LDDMM} framework, which is the Riemannian metric for the group $G_V$ in \cref{eq:GV}.

The above approaches all avoid learning the entire deformation, instead they learn a deformation that belongs to a specific class of deformation models. 
This makes it possible to embed the learned deformation model in a variational model for image reconstruction.

\subsection{Learning motion models}\label{sec:LearnedMotionModel}
The methods mentioned here deals with using deep learning in reconstruction with a motion model (\cref{sec:MotionRec}).
Many of the motion models are however sufficient for capturing the desired motion, so the main motivation with introducing deep learning is to speed up these methods. 

In particular, the above means we still aim to solve the penalised variational formulation \eqref{eqn:penalJointModel} with an explicit temporal model, such as the continuity equation \eqref{eqn:contEq}. The network then essentially learns to produce the motion field $\vfield(t,\cdot)$ from the time series $\signal(t,\cdot)$. Such a network can then be utilised to estimate the motion field, instead of solving the corresponding sub-problem \eqref{eqn:jointMotion} in the alternating minimisation. For instance, one could use neural networks that are designed to compute the optical flow \cite{dosovitskiy2015flownet,ilg2017flownet}.

Another possibility is to account for the explicit structure of the \ac{PDE} by using networks that aim to find a \ac{PDE} representation for given data \cite{long2019pde}.
Alternatively, one may build network architectures based on the discretisation of the underlying equations as motivated in \cite{arridge2019networks}.
Finally, similar to the work of joint motion estimation and reconstruction, one can learn a motion map that is used in a learned reconstructions scheme \cite{qin2018joint}.

\section{Outlook and conclusions}
The variational approaches outlined in \cref{sec:MotionRec,sec:DeforTemp}, and then in more detail in \cref{sec:ODE,sec:PDE}, rely on explicit parametrised temporal models. 
These temporal models are either given by deformation operators with time dependent parameters (\cref{sec:DeforTemp}) or through a motion model (\cref{sec:MotionRec}).
Powerful techniques from analysis and differential geometry can be used to characterise regularising properties of these reconstruction methods.
They also provide state-of-the-art results when applied to challenging tomographic data that is highly noisy and/or incomplete.
The methods are however difficult to use due to the computational burden and the sheer number of (regularisation) parameters that needs to be choosen. 

Data driven temporal modelling offers a way to address the computational burden inherent in the variational approaches.
Here, it is clear that deep learning needs to be embedded in such a way that the resulting learned temporal model is parametrised. VoxelMorph \cite{Balakrishnan:2019aa} and Quicksilver \cite{yang2017quicksilver} are examples of how this can be done in the context of diffeomorphic deformation, and \cite{Liu:2019aa,Pouchol:2019aa} show how such learned models can be used in reconstruction. 
In the near future, we expect more development along these lines.
Finding appropriate training data however remains a key difficulty in data driven approaches as in most dynamic imaging scenarios, there is no underlying ground-truth data available. Thus, most likely one will need to resort to simulations for training these models. Possibly, one could utilise reconstructions generated by variational approaches from experimental data as gold-standard reference reconstructions for a training procedure.
In conclusion, there is a great need for dynamic digital phantoms that include both natural image and motion features, that can serve as input for simulators.  

A final challenge that applies to all reconstruction methods in dynamic inverse problems is to formulate relevant validation and comparison protocols.

\bibliographystyle{spmpsci}
\bibliography{dyntomo_refs}

\end{document}